\UseRawInputEncoding
\documentclass[aps,pra,twocolumn,superscriptaddress,groupedaddress]{revtex4}

\usepackage{amssymb} \usepackage{color,graphicx} \usepackage{amsmath}
\usepackage{amsbsy} \usepackage{amsthm} \usepackage{bbm}
\usepackage{bm,bbm} \usepackage{float} \usepackage{braket}
\usepackage{placeins}
\usepackage[colorlinks=true,citecolor=blue,linkcolor=red,urlcolor=red]{hyperref}
\usepackage[sort&compress]{natbib}
\usepackage{comment}
\usepackage{stackengine}
\usepackage{hhline}

\newcommand{\bq}{\begin{equation}} \newcommand{\eq}{\end{equation}}
\newcommand{\bqali}{\bq\begin{aligned}}
\newcommand{\eqali}{\end{aligned}\eq}
\newcommand{\bqn}{\begin{equation*}}
\newcommand{\eqn}{\end{equation*}}
\newcommand{\TR}[2]{\text{Tr}^{(#1)}\left[#2\right]}

\newcommand\D{\operatorname{d}}
\renewcommand\k{{\bf k}}

\newcommand\Pp{{\bf P}}
\newcommand\Ll{{\bf L}}

\renewcommand\k{{\bf k}}
\renewcommand\r{{\bf r}}
\newcommand\p{{\bf p}}

\newcommand\x{{\bf x}}
\newcommand\z{{\bf z}}
\newcommand\kb{k_\text{\tiny B}}

\newcommand{\OO}{{\bm \Omega}}

\graphicspath{{../Figures/}}

\begin{document}
\title{A perturbative algorithm for rotational decoherence}

\author{Matteo Carlesso}
\email{matteo.carlesso@ts.infn.it}
\affiliation{Department of Physics, University of Trieste, Strada Costiera 11, 34151 Trieste, Italy}
\affiliation{Istituto Nazionale di Fisica Nucleare, Trieste Section, Via Valerio 2, 34127 Trieste, Italy}
\author{Hamid Reza Naeij}
\affiliation{Department of Chemistry, Sharif University of Technology, Tehran, Iran}
\author{Angelo Bassi}
\affiliation{Department of Physics, University of Trieste, Strada Costiera 11, 34151 Trieste, Italy}
\affiliation{Istituto Nazionale di Fisica Nucleare, Trieste Section, Via Valerio 2, 34127 Trieste, Italy}

\date{\today}
\begin{abstract}

Recent advances in levitated optomechanics provide new perspectives for the use of rotational degrees of freedom for the development of quantum technologies as well as for testing fundamental physics. As for the translational case, their use, especially in the quantum regime, is limited by environmental noises, whose characterization is  fundamental in order to assess, control and minimize their effect, in particular decoherence. Here, we present a general perturbative approach to compute decoherence for a quantum system in a superposition of its rotational degrees of freedom. The specific cases of the dipole-dipole and quadrupole-quadrupole interactions are solved explicitly, and we show that the rotational degrees of freedom decohere on a time scale that {can be} longer than the translational one. 

\end{abstract}
%
\maketitle

\section{Introduction}

{Decoherence is an {unavoidable} feature of quantum systems and, ultimately, sets the limits to the applicability of quantum mechanics while moving towards the macroscopic realm. Various {research fields}, from quantum foundations to quantum technologies, {investigate quantum systems which} are inevitably influenced and disturbed by the environmental decoherence.}
One of the main {experimental} challenges is to control, if not remove,   {their effects on}  quantum systems. The literature on this subject is vast;
environmental influences on the translational degrees of freedom of a material system \cite{Breuer:2002aa,Schlosshauer:2007aa} were  studied within the context of scattering theory \cite{Joos:1985aa,Gallis:1990aa,Hornberger:2003aa,Hornberger:2007aa,Gasbarri:2015aa} and Brownian motion \cite{Caldeira:1983aa,Hu:1992aa,Ferialdi:2017aa,Carlesso:2017aa}, with important applications to molecular interferometry \cite{Hornberger:2004aa,Bateman:2014aa,Belenchia:2019aa}, cold atoms \cite{Kovachy:2015aa,Becker:2018aa} and optomechanics \cite{Aspelmeyer:2014aa,Abbott:2016aa,Armano:2016aa,Vovrosh:2017aa,Vinante:2017aa,Hempston:2017aa,Vinante:2019aa}. See  \cite{Gardiner:2004aa,Clerk:2010aa} for a general review.

{Although one can try to minimize such decoherence effects, for example by developing experiments at low temperatures in ultra-high vacuum, the identification of alternative paths for longer decoherence time scales would be a game-changer. The avenue of levitated systems opens vast possibilities in this respect. Among them, the exploitation of rotational degrees of freedom could be a suitable solution for extending the decoherence time and thus enabling various possible applications of quantum mechanics to more macroscopic regimes than before.}
Recently, {the interest of the community has been triggered}
 for the new possibilities they offer both for research \cite{Paterson:2001aa,Bonin:2002aa,Shelton:2005aa,Jones:2009aa,Tong:2010aa,Arita:2013aa,Kuhn:2015aa,Hoang:2016ab,Kuhn:2017ab,Rashid:2018aa,Carlesso:2017ac,Schrinski:2017aa,Carlesso:2018ab,Stickler:2018aa}, as well as for technological applications     \cite{Bhattacharya:2007aa,Trojek:2012aa,Bhattacharya:2015aa,Shi:2016aa,Stickler:2016ab}. With that comes the necessity to understand and characterize  rotational decoherence. A first master equation was derived in \cite{Zhong:2016aa}, and then extended to include translational effects \cite{Stickler:2016aa} and small anisotropies of the system \cite{Papendell:2017aa}. 

In this {paper}, we approach from a more algorithmic point of view the problem of quantifying  decoherence effects on a quantum system initially prepared in a superposition of rotational degrees of freedom. First, we introduce a
general and exact expression for the decoherence rate, which  can be applied to every interaction potential between the system and its surrounding environment, and develop a perturbative expansion. 
Then, we provide the explicit and exact form of the rotational decoherence rate due to a dipole-dipole and quadrupole-quadrupole interaction. The corresponding results will show that rotations can be far less affected by decoherence than translations  \cite{Zhong:2016aa}. This means the rotational superpositions can be used to reach longer coherence times for fundamental studies, as well as for technology development.

\section{The model}

We consider a non-spherical particle of mass $M$, and we focus only on its rotational motion. Its orientation is described quantum mechanically in terms of the state $\ket \OO$, representing the system in the angular configuration $\OO$; This can be obtained by starting from a reference configuration $\ket{{\bm 0}}$ (e.g.~with the anisotropy of the system along the $x$ axis) then applying a rotation $\hat D_\text{\tiny S}(\OO)=e^{-i\hat L_z\gamma/\hbar}e^{-i\hat L_x\beta/\hbar}e^{-i\hat L_z\alpha/\hbar}$ defined by the three Euler angles $ \OO=({\alpha,\beta,\gamma})$ with $\hat L_i$ representing the angular momentum operator along the $i$-th axes \cite{Fischer:2013aa,Zhong:2016aa,Sakurai:2011aa}. 
The statistical operator describing the rotational state of the system is
\bq
\hat \rho_\text{\tiny S}=\int \D\OO\int \D\OO'\,{\rho}_\text{\tiny S}(\OO,\OO')\ket \OO \bra{\OO'},
\eq
where $\rho_\text{\tiny S}(\OO, \OO')$ are its matrix elements with respect to $\ket{\OO}$ and $\ket {\OO'}$. 
The system eventually couples to the surrounding environment.
The corresponding dynamics is described by 
the following master equation \cite{Zhong:2016aa}
\bq\label{mastergeneral}
\frac{\D \rho_\text{\tiny S}(\alpha,\alpha',t)}{\D t}=-\Lambda_\text{\tiny R}\rho_\text{\tiny S}(\alpha,\alpha',t).
\eq
where, to be quantitative, we considered the case where the system is in a superposition of angular configurations obtained only from rotations around the $z$ axis. 
Here, $\Lambda_\text{\tiny R}$ is the decoherence rate due to the $N$ particles of which the environment is made. The latter is defined as
\bq\label{expr5}
\Lambda_\text{\tiny R}=n\!\int\D k\, v(k) \rho(k)\frac{\int\D\hat\k'\int \D \hat\p'}{8\pi}|\Delta f^\omega(k\hat\k',k\hat\p')|^2,
\eq
with $n=N/V$ the number density, $v(k)=\hbar k/m_\text{\tiny gas}$ the velocity of environment particle of mass $m_\text{\tiny gas}$, $\rho(k)=4\pi k^2\mu(k)$  the momentum distribution of the environmental particles and 
\bq
\Delta f^\omega(\k',\p')=f(\k',\p')-f^\omega(\k',\p'),
\eq 
where $f(\k',\p')$ is the scattering amplitude, where we defined $\omega=\alpha-\alpha'$.
Here, $f^\omega(\k',\p')$ has the same form of $f(\k',\p')$ but with $\k'$ and $\p'$ replaced by $\k'_\omega$ and $\p'_\omega$ respectively, which are the same vectors rotated by the angle $\omega$. A derivation of Eq.~\eqref{mastergeneral}, alternative to that in Ref.~\cite{Zhong:2016aa}, is reported in Appendix \ref{App.dermaster}.

\section{Rotations under the Born approximation}
To further investigate the properties of the decoherence rate $\Lambda_\text{\tiny R}$, we apply the Born approximation. In this case, $f^\omega(k\hat \k',k\hat \p')$ is expressed as
\cite{Sakurai:2011aa}: 
\bq\label{scatt1}
f^\omega(k\hat \k',k\hat\p')=-\frac{ m_\text{\tiny gas}}{2\pi\hbar^2}\int\D\r\,V(\r)\,e^{-i{k(\hat\k'_\omega-\hat\p'_\omega)\cdot\r}{}}.
\eq
where $V(\r)$ is the interaction potential between the system and the environmental particle.
Being interested in rotations, we decompose $e^{-i\k\cdot\r}$ and $V(\r)$ in spherical harmonics $Y_{l,m}(\hat r)$: $e^{-i\k\cdot\r}=4\pi\sum_{lm}(-i)^lj_l(kr)Y_{l,m}(\hat \r)Y^*_{l,m}(\hat \k)$, where $j_l(x)$ are the spherical Bessel function of the first kind, and 
\bq
\label{eq.general.potential}
V(\r)=\sum_{l''m''}d_{l'',m''}(r)Y_{l'',m''}(\hat \r),
\eq
where $d_{l'',m''}(r)$ denotes the radial part of the potential and $\hat \r=\r/r$.
One obtains:
\bqali\label{scattamp}
\Delta f^\omega(k\hat \k',k\hat \p')=-\frac{8\pi  m_\text{\tiny gas}}{\hbar^2}\sum_{lm}\sum_{l'm'}\sum_{l''m''}R_{l,l',l'',m''}(k)\\
\times Y^*_{l,m}(\hat \k')Y_{l',m'}(\hat \p')G_{l,m,l',m',l'',m''}(\omega),
\eqali
where
\bq\label{eq.def.R}
R_{l,l',l'',m''}( k)=\int_0^{+\infty}\D r\,r^2j_l( {k r})j_{l'}( {k r})d_{l'',m''}(r),
\eq contains the information on the radial part of the potential, and the angular part is encoded in
\bqali
\label{eq.def.G}
G_{l,m,l',m',l'',m''}(\omega)
=(-)^{m'}i^{l'-l}\left(1-e^{i\omega(m-m')}\right)\\
\times \sqrt{\tfrac{(2l+1)(2l'+1)(2l''+1)}{4\pi}}
\begin{pmatrix}
l&l'&l''\\
m&-m'&m''
\end{pmatrix}\!\!
\begin{pmatrix}
l&l'&l''\\
0&0&0
\end{pmatrix},
\eqali
with $\begin{pmatrix}
p_1&p_2&p_3\\
s_1&s_2&s_3
\end{pmatrix}$ denoting the Wigner 3-j symbol. The latter vanishes except when $s_1+s_2+s_3=0$ and the numbers ${p_j}$ satisfy a triangle inequality: $p_i\leq p_j+p_k$, with $i,j,k=1,2,3$ but different among them. In particular, due to the first Wigner 3-j symbol appearing in Eq.~\eqref{eq.def.G}, we have always $m'=m+m''$.
 Given  Eq.~\eqref{eq.def.R} and Eq.~\eqref{eq.def.G}, one has determined the terms of Eq.~\eqref{scattamp}. Once one takes the square modulus of the latter and integrates it as in Eq.~\eqref{expr5}, one arrives at the rotational decoherence rate.

Since the decomposition in Eq.~\eqref{eq.general.potential} is fully general and it can be applied to any potential, the expression in Eq.~\eqref{scattamp} for $\Delta f^\omega(k\hat \k',k\hat \p')$ under the Born approximation is general as well. In particular, it provides a perturbative technique to be used for potentials whose exact expressions cannot be computed. Indeed, one can approximate the potential $V(\r)$ to the first terms of the sum in Eq.~\eqref{eq.general.potential}:
\bq\label{potentialTaylor}
V(\r)\simeq d_{0,0}(r)Y_{0,0}(\hat \r)+\sum_{m''=-1}^1d_{1,m''}(r)Y_{1,m''}(\hat \r)+\dots,
\eq
and still obtain an analytical expression for $\Lambda_\text{\tiny R}$ as well as the corresponding decoherence rate $\Lambda_\text{\tiny T}$ for the translational case. {In general, there is no strict rule for the level of truncation of the series in Eq.~\eqref{potentialTaylor}. This depends on the particular potential $V(\r)$, on the desired level of approximation and on the specific of the original potential one aims at retaining.}
For each term, one obtains the corresponding $R_{l,l',l'',m''}(k)$ from Eq.~\eqref{eq.def.R}, which, together with Eq.~\eqref{eq.def.G}, determines Eq.~\eqref{scattamp}.
Once merged with Eq.~\eqref{expr5}, it provides a straightforward algorithm to compute the decoherence rate of a system in angular superposition. In what follows, we study the first contributions for $l''=0$, 1 and 2 with explicit cases, which show how the algorithm works.

\section{Spherical interaction}
Consider the simple case of an interaction exhibiting spherical symmetry. The potential is given only by the first term in Eq.~\eqref{potentialTaylor}. This means that $l''=0$ and $m''=0$ in the third sum in Eq.~\eqref{scattamp}, and, correspondingly, Eq.~\eqref{eq.def.G} gives that $G_{l,m,l',m',0,0}$ is proportional to
\bq
\begin{pmatrix}
l&l'&0\\
m&-m'&0
\end{pmatrix}\!\!
\begin{pmatrix}
l&l'&0\\
0&0&0
\end{pmatrix}=
\frac{(-1)^m}{(2l+1)}\delta_{l,l'}\delta_{m,m'}.
\eq
Due to the presence of $(1-e^{i\omega(m-m')})$ in Eq.~\eqref{eq.def.G}, it follows that all $G_{l,m,l',m',0,0}$ are zero, leading to
$\Lambda_\text{\tiny R}=0$.

This result is not unexpected. Indeed, the symmetry of the interaction potential makes $f(\k'_\omega,\p'_\omega)$ independent from $\omega$, giving $\Delta f^\omega(\k',\p')=0$. Physically, the symmetrical interaction between the system and its environment means that the system interacts effectively as it was spherical, even if it is not.

\section{Dipole-dipole interaction}
The first non-zero contributions to $\Lambda_\text{\tiny R}$ are given by the second term in Eq.~\eqref{potentialTaylor}. These are three contributions with $l''=1$ and $m''=-1,0,1$. Thus,  in Eq.~\eqref{scattamp}, one can substitute $\sum_{l'',m''}\mathcal S_{l,m,l',m',l'',m''}$ with $\sum_{m''=-1}^1\mathcal S_{l,m,l',m',1,m''}$, where $\mathcal S$ denotes the terms of the sum. For each $m''$, one determines the corresponding $G_{l,m,l',m',1,m''}(\omega)$ through Eq.~\eqref{eq.def.G}. Due to the structure of the Wigner 3-j symbols, they are non vanishing only when $l'=l\pm1$ and $m'=m+m''$, thus imposing $\sum_{l'm'}\mathcal S_{l,m,l',m',1,m''}\to\sum_{s=-1,1}\mathcal S_{l,m,l+s,m+m'',1,m''}\theta_{l+s}$ in Eq.~\eqref{scattamp}, where $\theta_{x\geq0}=1$ and $\theta_{x<0}=0$.

In this case, Eq.~\eqref{scattamp}  strongly simplifies: $\sum_{l'=0}^{+\infty}\sum_{m'=-l'}^{l'}\sum_{l''=0}^{+\infty}\sum_{m''=-l''}^{l''}\mathcal S_{l,m,l',m',l'',m''}
$ reduces to $\sum_{s=-1,1}\sum_{m''=-1}^1\mathcal S_{l,m,l+s,m+m'',1,m''}\theta_{l+s}$. By merging these results with Eq.~\eqref{scattamp}, taking its square modulus and performing the angular integration we find
\begin{widetext}
\bq\label{scattamp2}
\int\D\hat\k'\int \D \hat\p'|\Delta f^\omega(k\hat\k',k\hat\p')|^2
=\frac{64\pi^2 m^2_\text{\tiny gas}}{\hbar^4}\sum_{lm}\sum_{s=-1,1}\sum_{m''=-1}^1
|R_{l,l+s,1,m''}(k)G_{l,m,l+s,m+m'',1,m''}(\omega)|^2,
\eq
\end{widetext}
where we took into account that $\set{Y_{l+s,m+m''}(\hat \p')}$ and $\set{Y_{l,m}^*(\hat \k')}$ are two sets of orthonormal functions. 

The corresponding coefficients $R_{l,l+s,1,m''}(k)$ depend on the radial behaviour of $d_{1,m''}(r)$ according to Eq.~\eqref{eq.def.R}. As an explicit example, we consider the interaction of a magnetic dipole with an environment made of magnetic dipoles, whose form reads \cite{Kranendonk:1963aa}
\bq
V(r,\hat\r_1,\hat\r_2)=\frac{\mu_0}{4\pi}\frac{\gamma_1\gamma_2}{r^3}\sum_{m''=-1}^1a_{m''} Y^*_{1,m''}(\hat \r_1)Y_{1,m''}(\hat \r_2),
\eq
where $\gamma_i$ is the modulus of the $i$-th dipole moment, $a_{\pm1}=1$, $a_0=-2$, and $r$ is the distance between the two dipoles. Here, $\hat\r_1$ identifies the orientation of the system, while $\hat\r_2$ that of the environmental dipole. The generalization to other dipole-dipole interactions is straightforward.
Such a potential can be expressed as in Eq.~\eqref{eq.general.potential}, with the only contributions given by $d_{1,m''}(r)={\tilde d_{1,m''}}/{r^3}$ with $m''=-1,0,1$ and
\bq\label{d-dipole}
{\tilde d_{1,m''}}=(-)^{m''+1}\frac{\mu_0}{4\pi }a_{m''}\gamma_1\gamma_2 Y^*_{1,m''}(\hat \r_1).
\eq
Correspondigly, Eq.~\eqref{eq.def.R} converges for $2l+s>0$ giving 
\bq\label{Rdipole}
R_{l,l+s,1,m''}(k)= \frac{2\sin(\tfrac\pi2s)\tilde d_{1,m''}}{\pi s(2l+s)}.
\eq
Once $R_{l,l+s,1,m''}(k)$ and $G_{l,m,l+s,m+m'',1,m''}(\omega)$   are determined, see Appendix \ref{appb}, we substitute them in Eq.~\eqref{scattamp2} and arrive at the result:
\begin{widetext}
\bq
\int\D\hat\k'\int \D \hat\p'|\Delta f^\omega(k\hat\k',k\hat\p')|^2=\frac{768}{\pi\hbar^4} m_\text{\tiny gas}^2\sin^2(\tfrac\omega2)(|\tilde d_{1,1}|^2+|\tilde d_{1,-1}|^2)\Sigma,
\eq
\end{widetext}
where $
\Sigma
\simeq2.16
$.
By 
averaging the above expression over the possible angular configurations of the system, we find
\bq\label{rotintegrated}
\braket{\int\D\hat\k'\int \D \hat\p'|\Delta f^\omega(k\hat\k',k\hat\p')|^2}=\frac{ m_\text{\tiny gas}^2 \mu_0^2\gamma_1^2\gamma_2^2}{\alpha\hbar^4}\sin^2(\tfrac\omega2),
\eq
with $\alpha\simeq0.15$, and we used Eq.~\eqref{d-dipole}.

By assuming that the environment is in thermal equilibrium at the temperature $T$, $\mu(k)$ is given by the Maxwell-Boltzmann distribution \cite{Schlosshauer:2007aa}
\bq\label{expr7}
\mu(k)=\left(\frac{\hbar^2}{2\pi m_\text{\tiny gas}\kb T}\right)^{3/2} \exp\left(-\frac{\hbar^2 k^2}{2m_\text{\tiny gas}\kb T}\right).
\eq 
Thus, by substituting $\rho(k)=4\pi k^2\mu(k)$ in Eq.~\eqref{expr5}, we get
\bqali
\Lambda_\text{\tiny R}=\frac{ m_\text{\tiny gas}^{3/2}}{(2\pi)^{3/2}\alpha\hbar^4} \mu_0^2\gamma_1^2\gamma_2^2n\sqrt{\kb T}\sin^2(\tfrac\omega2).
\eqali
This is an exact result.
The rotational decoherence rate $\Lambda_\text{\tiny R}$ depends on $\sqrt{T}$ and the angular superposition distance through $\sin^2(\omega/2)$: it vanishes for $\omega=0$ and it is maximum for $\omega=\pi$, which are respectively the cases of an aligned and anti-aligned superposition.

\subsection{Dipole-dipole interaction: translational case}
\begin{figure}[t!]
\centering
\includegraphics[width=\linewidth]{{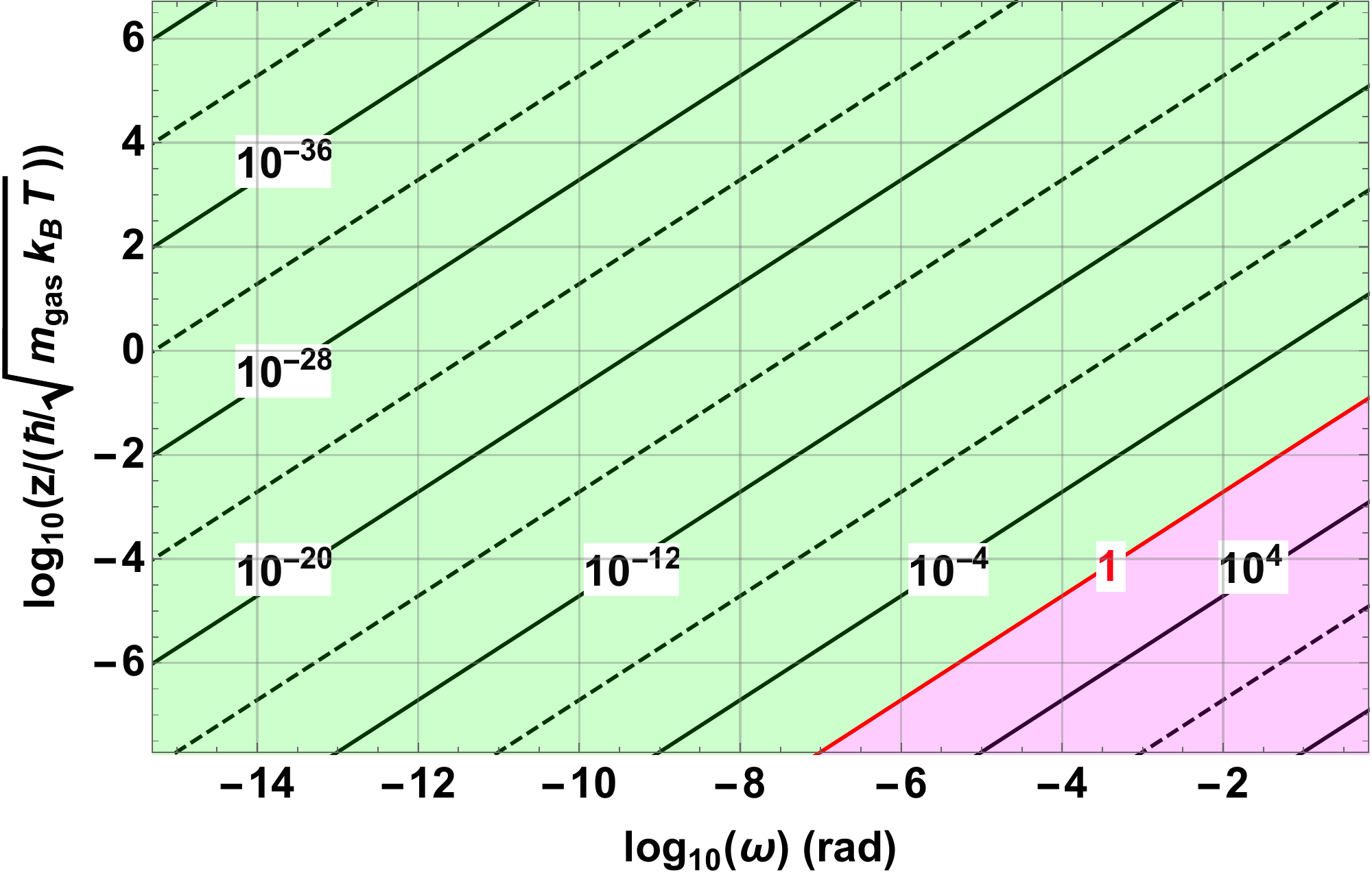}}
\includegraphics[width=\linewidth]{{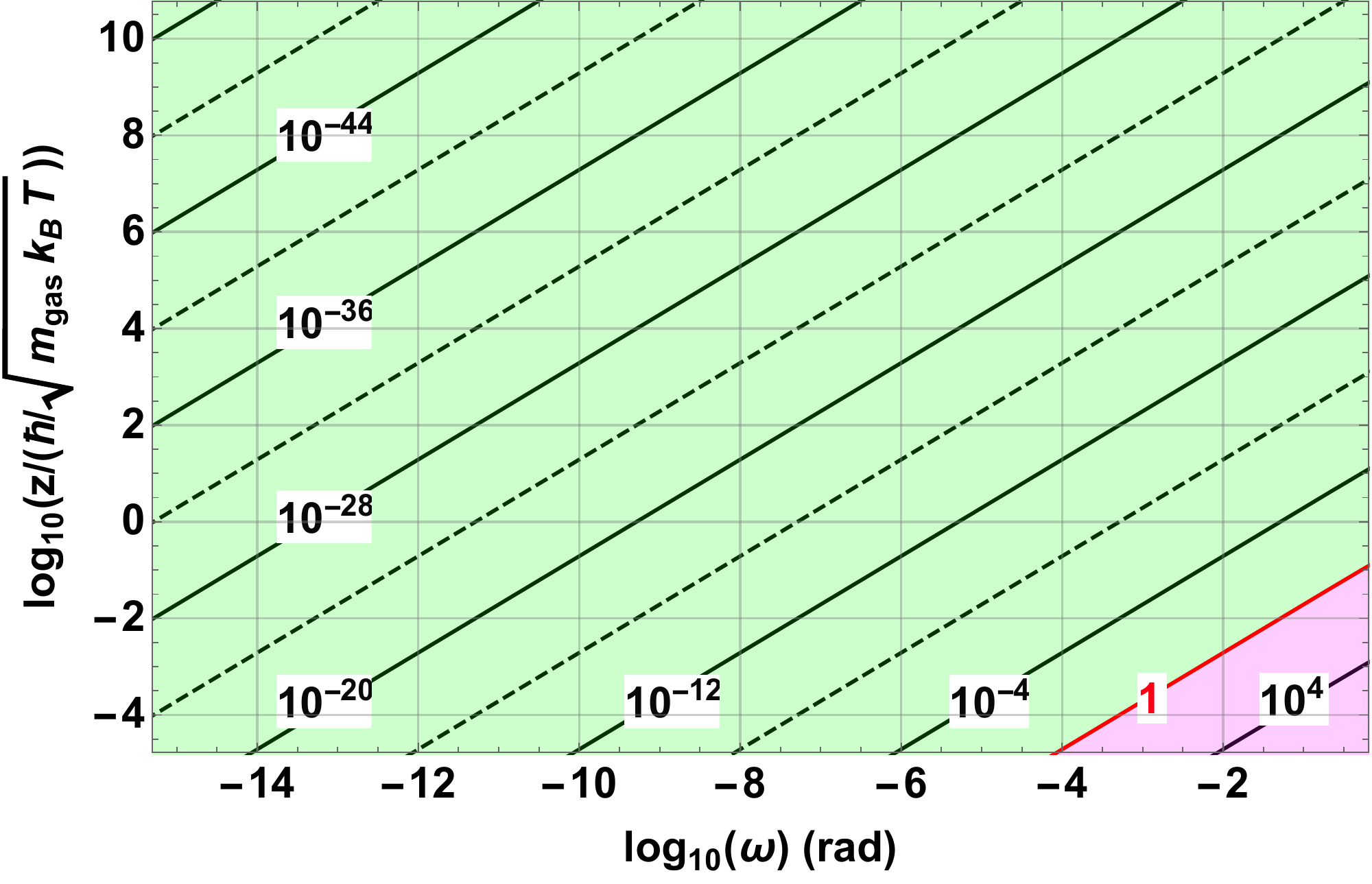}}
\caption{\label{figrapLambda} Comparison between the rotational and translational decoherence rates for the dipole-dipole interaction. The level curves of $\Lambda_\text{\tiny R}/\Lambda_\text{\tiny T}$ are reported as functions of the superposition angle $\omega$ and {the adimensional superposition distance $z/(\hbar/\sqrt{m_\text{\tiny gas}\kb T})$. The latter corresponds, for both the panels, to a superposition distance $z$ ranging from $\sim0.3$\,m to $\sim 10^{-15}$\,m.} Here we took as reference values $T=10^{-4}\,$K (top panel) and $T=100\,$K (bottom panel). The gas mass is taken equal to $m_\text{\tiny gas}\simeq10^{-26}\,$kg ($\sim$ light molecule). }
\end{figure}
For a comparison, we compute the translational decoherence rate for a dipole-dipole interaction. 
The translational rate can be obtained applying the following substitution in Eq.~\eqref{expr5}:
\bq\label{rottotran}
\tfrac12|\Delta f^\omega(k\hat\k',k\hat\p')|^2\to|f(k\hat \k',k\hat\p')|^2\left(1-e^{ik{(\hat\k-\hat\p')\cdot(\x-\x')}{}}\right).
\eq
Following a similar strategy as described for the rotational case, in the short length limit ($kz\ll1$ with $\z=(\x-\x')$ and $z=|\z |$) we obtain
\bqali\label{integratedtrans}
&\braket{\int\D\hat\k'\int \D \hat\p' | f(k\hat\k',k\hat\p')|^2(1-e^{ik(\hat\k-\hat\p')\cdot(\x-\x')})}\\
&=\frac{m^2_\text{\tiny gas}\mu_0^2\gamma_1^2\gamma_2^2}{\alpha_1\hbar^4} k^2z^2,
\eqali
with $\alpha_1\simeq0.18$. We thus find the translational decoherence rate
\bqali\label{LambdaTr}
\Lambda_\text{\tiny T}=\left(\frac{2}{\pi}\right)^{3/2}\frac{ m_\text{\tiny gas}^{5/2}}{\alpha_1\hbar^6} \mu_0^2\gamma_1^2\gamma_2^2n({\kb T})^{3/2}z^2,
\eqali
which is proportional to $T^{3/2}$ and depends on $|\x-\x'|^2$.

We can now compute the ratio of the two decoherence rates, which reads
\bq\label{ratio}
\Lambda_\text{\tiny R}/\Lambda_\text{\tiny T}=\frac{\alpha_1}{\alpha}\frac{\hbar^2}{8m_\text{\tiny gas}\kb T}\frac{\sin^2(\omega/2)}{z^2}.
\eq
The ratio depends on the superposition distances as ${\sin^2(\omega/2)}/{z^2}$, but most importantly, it scales with the inverse temperature.
Figure \ref{figrapLambda} shows the ratio of the two decoherence rates for two values of the temperature of the environment ($10^{-4}\,$K and $100\,$K) with varying $\omega$ and $z$. As it is clear from Figure \ref{figrapLambda}, the ratio €‹decreases by increasing the temperature of the environment, while both rates increase with the temperature. Moreover, one can conclude that for a given temperature, the rotational decoherence time ($1/\Lambda_\text{\tiny R}$) can be much longer than the translational decoherence time ($1/\Lambda_\text{\tiny T}$). This corresponds to the green region in Fig.~\ref{figrapLambda}.

\section{Quadrupole-quadrupole interaction}
 The next order contribution to $\Lambda_\text{\tiny R}$ is given by $l''=2$ and corresponding $m''=-2,-1,0,1,2$. 
Also in this case, Eq.~\eqref{eq.def.G} is such that only the terms with $l+l'\geq2$ and $|l-l'|\leq2$ give non-vanishing contributions. 
As an explicit example, we consider a modified anisotropic intermolecular interaction of the quadrupole-quadrupole form, whose corresponding potential is of the form
\bq\label{potential4}
V(r,\hat \r_1,\hat \r_2)=\frac{4\pi \mu_1\mu_2}{r^4}\sum_{m''=-2}^2a_{m''}Y_{2,m''}(\hat \r_1)Y^*_{2,m''}(\hat \r_2),
\eq
where $r$ is the distance among the molecules, $\hat \r_1$ and $\hat \r_2$ identify respectively the orientation of the system and the environmental particle, 
 $\mu_i$ quantify the quadrupole momentum of the molecules, $a_{\pm 2}=1$, $a_{\pm 1}=-4$ and $a_0=6$ \cite{Kranendonk:1963aa}.
By comparing the latter expression with Eq.~\eqref{eq.general.potential}, we can define $d_{2,m''}(r)=\tilde d_{2,m''}/r^4$, with
\bq
\tilde d_{2,m''}=4\pi \mu_1\mu_2a_{m''}Y^*_{2,m''}(\hat \r_2).
\eq 
{The potential in Eq.~\eqref{potential4} is a modification of the typically used quadrupole-quadrupole potential which scales with $r^{-5}$ \cite{Kranendonk:1963aa}. For the sake of simplicity, we consider the form in Eq.~\eqref{potential4} to avoid  short-length divergences. Our method can be safely applied also for the $\sim1/r^5$ potential by considering a suitable short-length cut-off when doing the integral in Eq.~\eqref{eq.def.R}.}
 With this choice, we find that Eq.~\eqref{eq.def.R} converges for $2l+s>1$ giving 
 \bq
 R_{l,l+s,2,m''}(k)=\frac{4\tilde d_{2,m''}k \cos(\tfrac\pi2 s)}{\pi(1-s^2)(s^2+4ls+4l^2-1)},
 \eq
 where $s=l'-l$.
 The explicit form of the only non-vanishing terms $G_{l,m,l',m',2,m''}$ and corresponding $R_{l,l',2,m''}$ are reported in Appendix \ref{appb}.

\begin{figure}[t!]
\centering
\includegraphics[width=\linewidth]{{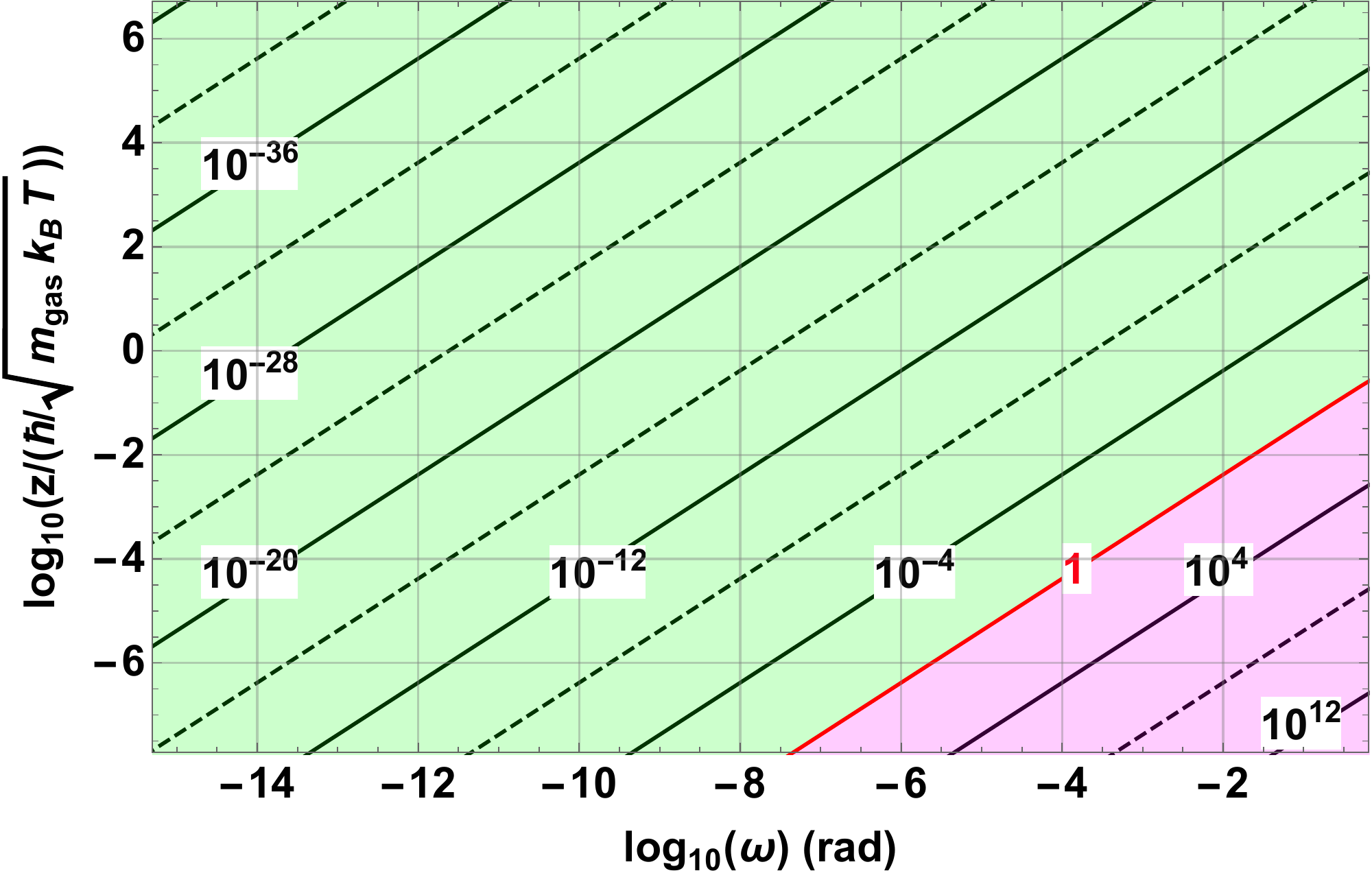}}
\includegraphics[width=\linewidth]{{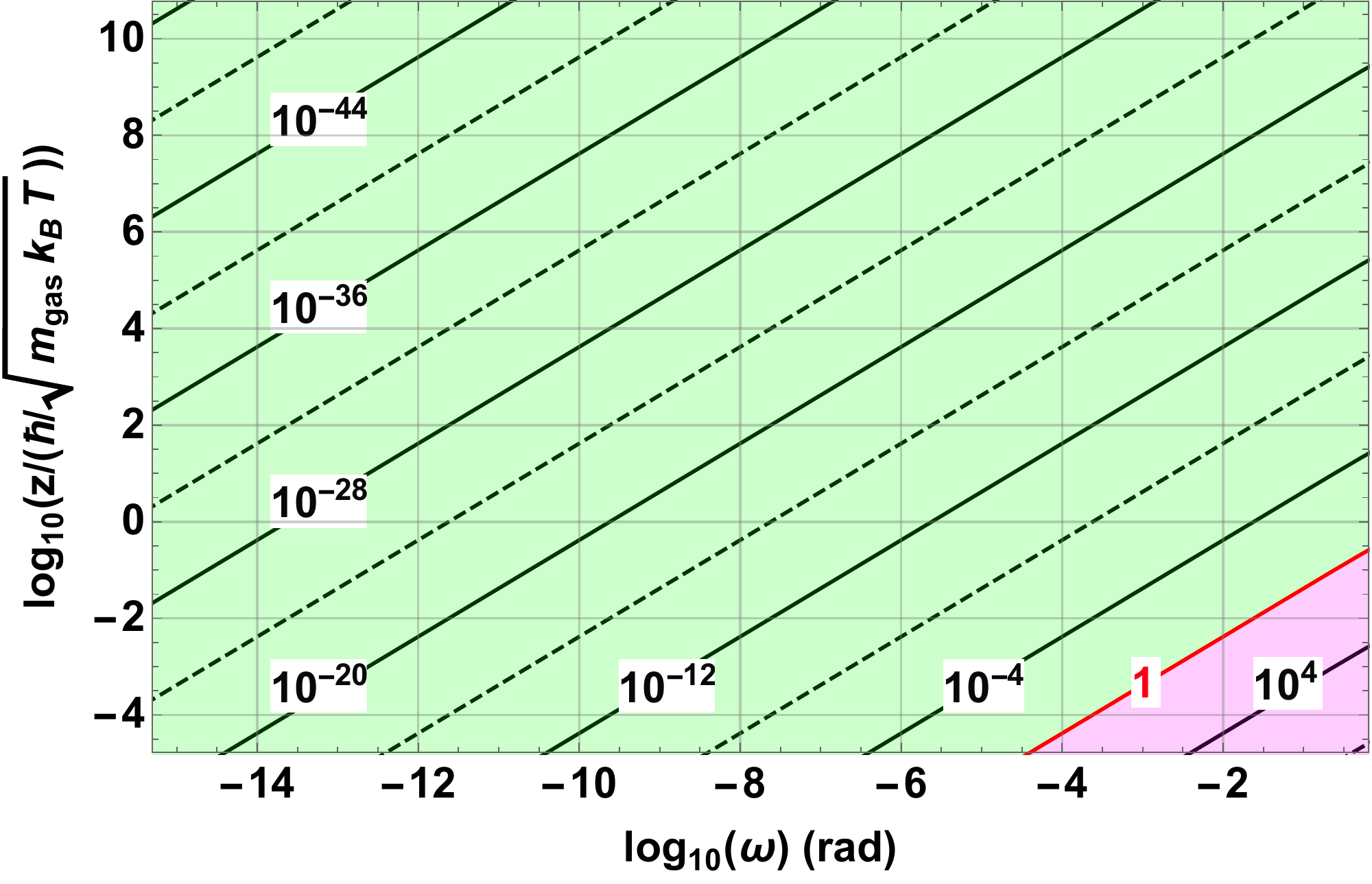}}
\caption{\label{fig4pole} Comparison of the rotational and translational decoherence rates for the quadrupole-quadrupole interaction. The level curves of $\Lambda_\text{\tiny R}/\Lambda_\text{\tiny T}$ are reported as functions of the superposition angle $\omega$ and {the adimensional superposition distance $z/(\hbar/\sqrt{m_\text{\tiny gas}\kb T})$. The latter corresponds, for both the panels, to a superposition distance $z$ ranging from $\sim0.3$\,m to $\sim 10^{-15}$\,m.} Here we took as reference values $T=10^{-4}\,$K (top panel) and $T=100\,$K (bottom panel). The gas mass is taken as $m_\text{\tiny gas}\simeq10^{-26}\,$kg ($\sim$ light molecule). }
\end{figure}

By following the procedure delineated above, one finds
\begin{widetext}
\bq
\braket{\int\D\hat\k'\int \D \hat\p'|\Delta f^\omega(k\hat\k',k\hat\p')|^2}
=\frac{ m^2_\text{\tiny gas}\mu_1^2\mu_2^2k^2}{\hbar^4}\left[\beta_1 \sin^2(\tfrac\omega2)+\beta_2 \sin^2(\omega)	\right],
\eq
\end{widetext}
where $\beta_1\simeq1.65 \times 10^5$ and $\beta_2\simeq1.75 \times 10^4$. Finally, by merging with Eq.~\eqref{expr5} and integrating over the Boltzmann distribution [cf.~Eq.~\eqref{expr7}], we have
\bq
\Lambda_\text{\tiny R}=\frac{\sqrt{2} \mu_1^2 \mu_2^2 m_\text{\tiny gas}^{5/2} ({\kb} {T})^{3/2}n}{\pi ^{3/2} \hbar ^6}\left[\beta_1 \sin^2(\tfrac\omega2)+\beta_2 \sin^2(\omega)	\right].
\eq
We can compare again the latter expression with the corresponding translational one, which can be derived similarly to Eq.~\eqref{LambdaTr} and reads
\bq
\Lambda_\text{\tiny T}=\frac{\beta_\text{\tiny T} \mu_1^2 \mu_2^2 m_\text{\tiny gas}^{7/2} z^2 (\kb T)^{5/2}n}{\hbar ^8},
\eq
where $\beta_\text{\tiny T} \simeq8.6 \times 10^4$. Thus, one obtains the ratio
\bq
\Lambda_\text{\tiny R}/\Lambda_\text{\tiny T}=\frac{\sqrt{2}\hbar^2}{\pi^{3/2}m_\text{\tiny gas}\kb T}\frac{\beta_1 \sin^2(\tfrac\omega2)+\beta_2 \sin^2(\omega)}{z^2\beta_\text{\tiny T}},
\eq
which we study in Fig.~\ref{fig4pole} for two different temperatures. As for the case of a dipole-dipole interaction, we see that rotational degrees of freedom {can be} less influenced by decoherence than translational ones for a large choice of the parameters.

\section{Conclusions}

In this work, we proposed a perturbative algorithm to quantify the environmental decoherence effects on a quantum system prepared in a superposition of rotational degrees of freedom. We showed that our  approach can be suitably applied to any potential that can be conveniently expressed in terms of spherical harmonics. For instance, we studied explicitly the examples of the dipole-dipole and quadrupole-quadrupole interactions. We obtained the explicit form of the rotational decoherence rate for such systems. By applying the same approach, we also evaluated the corresponding translational decoherence rates and performed a comparison among the two. It turns out that -- for both examples -- rotational degrees of freedom {can be} less influenced by decoherence than translational ones {for a large choice of the parameter space of superposition angles and distances, which is highlighted by the green region in Figure \ref{figrapLambda} and Figure \ref{fig4pole}}.  Qualitatively similar results were found in \cite{Zhong:2016aa}, where {the simple case of a Gaussian potential elongated along one direction}
was considered {to solve explicitly Eq.~\eqref{scatt1}}. The advantage of working with rotational degrees of freedom is even stronger when moving to higher temperatures, as it is reported in Figure \ref{figrapLambda} and Figure \ref{fig4pole}. 
Thus, our approach can be of relevance for the calibration of decoherence effects also beyond what is usually considered the quantum realm.

\section*{Acknowledgments}
MC and AB acknowledge financial support from the H2020 FET Project TEQ (Grant No. 766900). HRN acknowledges financial support from the Ministry of Science, Research and Technology of IRAN, and hospitality from the University of Trieste, where part of this work was carried out. AB acknowledges financial support from the COST Action QTSpace (CA15220) and INFN. This research was supported by grant number (FQXi-RFP-CPW-2002) from the Foundational Questions Institute and Fetzer Franklin Fund, a donor advised fund of Silicon Valley Community Foundation.

   \onecolumngrid
\appendix

\section{Derivation of the master equation}
\label{App.dermaster}

We assume that at time $t=0$ the system and the environment are decoupled, and the total initial state is $\hat \rho_\text{\tiny T}=\hat \rho_\text{\tiny S}\otimes\hat \rho_\text{\tiny E}$, where $\hat\rho_\text{\tiny E}$ is the state of the environment. 
Starting from the configuration ${\bm\Omega}$, a scattering process at time $t$ can be described as 
\bq
\ket{\OO}\otimes\ket{	\chi}		\quad\xrightarrow[]{\text{scattering}}\quad		\ket{ \OO}\otimes\hat{\mathbb S}_{\OO}\ket{\chi},
\eq
where $\ket\chi$ is a generic state of the environment and the recoil-less limit is considered ($M\gg m_\text{\tiny gas}$, where $m_\text{\tiny gas}$ is the mass of the environmental particle). In this case, the scattering operator $\hat{\mathbb S}_{\bm \Omega}$ acts on the environmental state only. $\hat{\mathbb S}_{\bm \Omega}$ can be related to the standard unitary scattering operator $\hat{\mathbb S}_{0}$ acting in  ${\bm \Omega}={\bm 0}$, through a rotation from the configuration $\ket{\bm 0}$ to $\ket{\bm \Omega}$: $\hat{\mathbb S}_{\bm \Omega}=\hat D_\text{\tiny E}({\bm \Omega})\hat{\mathbb S}_{0}\hat D_\text{\tiny E}^\dag({\bm \Omega})$.

At time $t$, after the scattering process has taken place, the system matrix elements change to ${\rho}_\text{\tiny S}(\OO,\OO') \eta({\bm \Omega},{\bm \Omega}')$,
where
\bq
\eta({\bm \Omega},{\bm \Omega}')=\TR{B}{\hat\rho_\text{\tiny E}\hat{\mathbb S}_{{\bm \Omega}'}^{\dag}\hat{\mathbb S}_{{\bm \Omega}}}.
\eq
To explicitly evaluate the partial trace over the degrees of freedom of the environment, we consider the total system to be confined in a box of volume $V$, and we assume the environment at thermal equilibrium, whose state is given by 
\bq
\hat \rho_\text{\tiny E}=\frac{(2\pi)^3}{V}\int\D k\,k^2 \mu(k)\sum_{l=0}^{+\infty}\sum_{m=-l}^l\ket{k,l,m}\bra{k,l,m},
\eq
where we assumed that the momentum distribution of the environmental particles is invariant under rotations, thus $\mu(\k)=\mu(k)$. Here, $\ket{k,l,m}$ is the common eigenstate of the momentum $\hat \Pp^2$, the total angular momentum $\hat \Ll^2$ and its $z$ component $\hat L_z$ of the environmental particle. In particular, the relation between the usual momentum eigenstate $\ket\p$ and $\ket{k,l,m}$ is given by \cite{Sakurai:2011aa}:
\bq\label{rel-pp}
\braket{\p|k,l,m}={\delta(p-k)}Y_{l,m}(\hat \p)/p,
\eq
where $Y_{l,m}(\hat \p)$ denotes the spherical harmonic and $\hat \p=\p/p$.

To be quantitative, let us consider the case where the system is in a superposition of angular configurations around the $z$ axis. This will be also the case of interest in most experimental setups, where one usually considers one direction per time. The state of the system can be then identified by $\ket\alpha=\hat D_\text{\tiny E}(\alpha)\ket{\bm 0}$, with $\hat D_\text{\tiny E}(\alpha)=\exp(-\tfrac i\hbar \hat L_z\alpha)$. Thus, we have
\bq\label{eqeta1}
\eta({\alpha},{ \alpha}')
=\frac{(2\pi)^3}{V}\int\D k\, k^2\mu(k)\sum_{l,m}\braket{k,l,m|\hat{\mathbb S}_{{\alpha}'}^{\dag}\hat{\mathbb S}_{{\alpha}}|k,l,m},
\eq
where the relation $\hat D_\text{\tiny E}(\alpha)\ket{k,l,m}=e^{-im\alpha}\ket{k,l,m}$ holds. We express the scattering matrix as
\bq
\hat{\mathbb S}_{{\alpha}}=\hat D_\text{\tiny E}(\alpha)(1+i\hat T)\hat D^\dag_\text{\tiny E}(\alpha),
\eq
 where $\hat T$ is the T-matrix of scattering theory \cite{Sakurai:2011aa}. Due to the unitarity of $\hat{\mathbb S}_{{\alpha}}$, one has that $-i(\hat T^\dag-\hat T)=\hat T^\dag\hat T$.
By exploiting these relations one finds
\bq
\eta({\alpha},{ \alpha}')=1-\frac{(2\pi)^3}{V}\int\D k\, k^2\mu(k)\sum_{lm}\int\D p\,p^2 \sum_{l'm'}\left(1-e^{-i(m-m')(\alpha-\alpha')}\right)\braket{k,l,m|\hat T^\dag|p,l',m'}\braket{p,l',m'|\hat T|k,l,m},
\eq
where the matrix elements of $\hat T$ can be expressed in the momentum space as \cite{Sakurai:2011aa}:
\bq
\braket{\k''|\hat T|\p''}
=-\delta(k''-p'')f(\k'',\p'')/(2\pi p''),
\eq
where $f(\k'',\p'')$ is the scattering amplitude. This, together with Eq.~\eqref{rel-pp}, brings to
\bq
\braket{p,l',m'|\hat T|k,l,m}=-\int\D\hat \k''\int\D\p''Y_{l,m}(\hat\k'') Y^*_{l',m'}(\hat \p'')f(k\hat\k'',k\hat\p'')\frac{\delta(p-k)}{2\pi}.
\eq
Consequently, one obtains
\bqali\label{expr3}
\eta({\alpha},{ \alpha}')=&1-\frac{t}{V}\int\D k\,\frac{\hbar k}{M}k^2 \mu(k)\sum_{lm}\sum_{l'm'}\left(1-e^{-i(m-m')(\alpha-\alpha')}\right)\\
&\int\D\hat\k'\int\D\hat\p'\int \D\hat\k''\int\D\hat\p''\,Y_{l,m}(\hat \k'')Y_{l,m}^*(\hat \k')
Y_{l',m'}(\hat \p')Y_{l',m'}^*(\hat \p'')f^*(k\hat\k',k\hat\p')f(k\hat\k'',k\hat\p''),
\eqali
where we exploited the normalization of the squared Dirac-$\delta$:
\bq
\left(\delta(p-k)\right)^2\sim\frac{\hbar p t}{2\pi M}\delta(p-k),
\eq
which is valid under the assumption that the decoherence time is larger than the collision time \cite{Schlosshauer:2007aa} -- usually considered instantaneous.

We now take into account that the spherical harmonics are of the form
\bq
Y_{l,m}(\hat\k)=(-)^m\sqrt{\frac{(2l+1)(l-m)!}{2\pi(l+m)!}}F_{l,m}(\theta_\k)e^{im\phi_\k},
\eq
where $(\theta_\k,\phi_\k)$ identify $\hat \k$ and $F_{lm}(\theta)$ is the Legendre polynomial. Consequently, we can rewrite
\bq
Y_{l,m}(\hat\k'')e^{-im\alpha}=Y_{l,m}(\hat\k''_{\alpha}),
\eq
where $\hat\k''_{\alpha}$ is obtained from $\hat\k''$ after a rotation $\alpha$ around $z$. Then, the phase in the parenthesis in Eq.~\eqref{expr3} can be absorbed in the spherical harmonics, and we find that
\bq
e^{-i(m-m')(\alpha-\alpha')}Y_{l,m}(\hat \k'')Y_{l,m}^*(\hat \k')
Y_{l',m'}(\hat \p')Y_{l',m'}^*(\hat \p'')
=Y_{l,m}(\hat \k''_\alpha)Y_{l,m}^*(\hat \k'_{\alpha'})
Y_{l',m'}(\hat \p'_{\alpha'})Y_{l',m'}^*(\hat \p''_\alpha).
\eq
Now, by exploiting the orthonormality of the spherical harmonics \cite{Zettili:2009aa}
\bq
\sum_{lm}Y_{l,m}(\hat \k)Y^*_{l,m}(\hat \k')=\delta(\hat \k-\hat \k'),
\eq
we obtain
\bq\label{expr4}
\eta(\alpha,\alpha')=1-\frac{t}{V}\int\D k\, \frac{\hbar k}{m_\text{\tiny gas}}k^2 \mu(k)\int\D\hat\k'\int \D \hat\p'f(k\hat \k',k \hat \p')\left(f^*(k\hat \k',k \hat \p')-f^*(k\hat \k'_\omega,k \hat \p'_\omega)\right).
\eq
The final result is reported in Eq.~\eqref{mastergeneral} of the main text, where one exploits the relation $\eta(\alpha,\alpha')=1-t\Lambda_\text{\tiny R}/N$, and $\Lambda_\text{\tiny R}$ is the $N$-particle decoherence rate reported in Eq.~\eqref{expr5} of the main text.
To derive the latter, we also considered that the state is self-adjoint $\hat \rho_\text{\tiny S}=\hat \rho_\text{\tiny S}^\dag$, which implies $\eta(\alpha,\alpha')=\eta^*(\alpha',\alpha)$.\\

\subsection{Comparison with the translational case}
It is worth to notice that the expression for $\Lambda_\text{\tiny R}$ in Eq.~\eqref{expr5} of the main text has the same formal structure of the master equation describing decoherence for the translational degrees of freedom \cite{Schlosshauer:2007aa}:
\bq\label{eqmastertrans}
\Lambda_\text{\tiny T}(\x,\x')=n\int\D k\,v(k)\rho(k)\frac{\int\D\hat \k'\int\D\hat \p'}{4\pi}|f(\k,k\hat\p')|^2\left(1-e^{i{(\k-k\hat\p')\cdot(\x-\x')}{}}\right).
\eq
Such an equation is obtained by replacing $\tfrac12|\Delta f^\omega(k\hat\k',k\hat\p')|^2$ in Eq.~(3) of the main text with
$f(k\hat\k',k\hat\p')\left(f^*(k\hat\k',k\hat\p')-{f^*(k\hat\k'_{\x-\x'},k\hat\p'_{\x-\x'})}{}\right)$.
Here $\k_{\x-\x'}$ is the vector $\k$ translated in space by the quantity $\x-\x'$. This result can be understood once we consider the 
 expression for the scattering amplitude generated by the potential $V(\r)$, under the Born approximation \cite{Zhong:2016aa}, which is given by Eq.~(5) of the main text with $\omega=0$,
and substitute to $\hat {\mathbb S}_\alpha$ in Eq.~\eqref{eqeta1} the scattering operator implementing the 
translation in space $\hat{\mathbb S}_{\x}=e^{-i{\hat \p\cdot\x}{}}\hat{\mathbb S}_0e^{i{\hat \p\cdot\x}{}}$. In this way, one obtains the expression in Eq.~\eqref{eqmastertrans}.

\section{Coefficients for the explicit examples}\label{appb}

Here we report the explicit form of the coefficients $G_{l,m,l',m',l'',m''}(\omega)$ and $R_{l,l',l'',m''}( k)$ which are respectively defined in Eq.~\eqref{eq.def.G} and Eq.~\eqref{eq.def.R}. For the dipole-dipole interaction they are reported in Table \ref{tabG1}, while for the quadrupole-quadrupole interaction in Table \ref{tabG2}.  

To derive the corresponding translational decoherence rate, one imposes the transformation in Eq.~\eqref{rottotran}. Here, both the scattering amplitude $f(k\hat \k',k\hat\p')$ as well as the angular dependance of the phase $e^{ik{(\hat\k-\hat\p')\cdot(\x-\x')}{}}$ -- in the short length limit  ($kz\ll1$ with $\z=(\x-\x')$ and $z=|\z |$) -- can be expressed in terms of spherical harmonics. In particular, one has that
\bq
\int\D\hat\k'\int \D \hat\p'| f(k\hat\k',k\hat\p')|^2(1-e^{ik(\hat\k-\hat\p')\cdot(\x-\x')})
=\frac{32\pi^2 m^2_\text{\tiny gas}k^2z^2}{\hbar^4}\sum_{lm}\sum_{l'm'}\sum_{l''m''}
|R_{l,l',l'',m''}(k)\tilde G_{l,m,l',m',l'',m''}|^2,
\eq
where  $R_{l,l',l'',m''}( k)$ keeps the same form as for the rotational degrees of freedom, while 
\bq
\tilde G_{l,m,l',m',l'',m''}
=(-)^{m'}i^{l'-l}
 \sqrt{\tfrac{(2l+1)(2l'+1)(2l''+1)}{4\pi}}
\begin{pmatrix}
l&l'&l''\\
m&-m'&m''
\end{pmatrix}\!\!
\begin{pmatrix}
l&l'&l''\\
0&0&0
\end{pmatrix},
\eq
differs from the definition in Eq.~\eqref{eq.def.G}.

\begin{table}[h!]
	\caption{\label{tabG1} Only non-vanishing terms $G_{l,m,l',m',1,m''}$ and corresponding values of $R_{l,l',1,m''}$ for the dipole-dipole interaction, as defined in Eq.~\eqref{eq.def.G} and Eq.~\eqref{Rdipole} respectively. For the rotational case we have only the contributions from $m''=+1$ and $m''=-1$, while for the translational case we also have the contribution from $m''=0$. Here we have: $l'=l+s$, $m'=m+m''$. Moreover, $\theta_{x>=0}=1$ and $\theta_{x<0}=0$.  
 }
\begin{tabular}{|c|c|c|c|c|}
\hline
$s$	&$m''$	&$G_{l,m,l+s,m+m'',1,m''}/\left(1-e^{i \omega }\right)$	&$R_{l,l+s,1,m''}/\tilde d_{1,m''}$\\		
\hline
\hline
-1	&-1		&$\frac{i}{2}  \sqrt{\frac{3(l+m-1) (l+m)}{2 \pi(4 l^2-1)}}\theta_{l+m-2}\theta_{l-1}$ & $\frac{2 }{\pi(2 l-1)}, \text{for } l>\tfrac12$\\
\hline
-1	&0		&$0$& $\frac{2 }{\pi(2 l-1)}, \text{for } l>\tfrac12$\\
\hline
-1	&1		&$-\frac{i}{2}    e^{-i \omega } \sqrt{\frac{3(l-m-1) (l-m)}{2 \pi(4 l^2-1)}}\theta_{l-m-2}\theta_{l-1}$&  $\frac{2 }{\pi(2 l-1)}, \text{for } l>\tfrac12$\\
\hline
1	&-1		&$\frac{i}{2}\sqrt{\frac{3(l-m+1) (l-m+2)}{2 \pi(4 l (l+2)+3)}}$& $\frac{2 }{\pi(2 l-1)}$\\
\hline
1	&0		&$0$&$\frac{2 }{\pi(2 l-1)}$\\
\hline
1	&1		&$-\frac{i}{2}   e^{-i \omega } \sqrt{\frac{3(l+m+1) (l+m+2)}{2 \pi(4 l (l+2)+3)}}$&$\frac{2 }{\pi(2 l-1)}$\\
\hline
\end{tabular}
\end{table}
 \begin{table*}[h!]
	\caption{\label{tabG2} {Only non-vanishing terms $G_{l,m,l',m',2,m''}$ and corresponding values of $R_{l,l',2,m''}$ for the quardupole-quardupole interaction, as defined in Eq.~\eqref{eq.def.G} and Eq.~\eqref{Rdipole} respectively. For the rotational case we have only the contributions from $m''=-2, -1,+1, +2$. Here we have: $l'=l+s$, $m'=m+m''$. Moreover, $\theta_{x>=0}=1$ and $\theta_{x<0}=0$. } 
 }
\begin{tabular}{|c|c|c|c|c|}
\hline
$s$	&$m''$	&$(-)^{l+1}G_{l,m,l+s,m+m'',2,m''}$	&$R_{l,l+s,2,m''}/(\tilde d_{2,m''}k)$\\		
\hline
\hline
-2	&-2		&$(1-e^{2i\omega})\frac{\sqrt{15} \sqrt[4]{\pi } 2^{l-6}}{\left(\frac{3}{2}-l\right)!} \sqrt{\frac{l! (l+m)!}{\left(2 l^3-3 l^2+l\right) \left(l+\frac{1}{2}\right)! (2 l-4)! (l+m-4)!}}\theta_{l+m-4}\theta_{l-2}$ & $\frac{4}{3\pi(4l^2-8l+3)}$\\
\hline
-2	&-1		&$-(1-e^{i \omega })\frac{\sqrt{15} \sqrt[4]{\pi } 2^{l-5}}{\left(\frac{3}{2}-l\right)!} \sqrt{\frac{l! (l-m)! (l+m)!}{\left(2 l^3-3 l^2+l\right) \left(l+\frac{1}{2}\right)! (2 l-4)! (l-m-1)! (l+m-3)!}}\theta_{l+m-3}\theta_{l-m-1}\theta_{l-2}$ & $\frac{4}{3\pi(4l^2-8l+3)}$\\
\hline
-2	&0		&$0$ & $\frac{4}{3\pi(4l^2-8l+3)}$\\
\hline
-2	&1		&$-(1-e^{-i\omega})\frac{\sqrt{15} \sqrt[4]{\pi } 2^{l-5}}{\left(\frac{3}{2}-l\right)!} \sqrt{\frac{l! (l-m)! (l+m)!}{\left(2 l^3-3 l^2+l\right) \left(l+\frac{1}{2}\right)! (2 l-4)! (l-m-3)! (l+m-1)!}}\theta_{l-m-3}\theta_{l+m-1}\theta_{l-2}$ & $\frac{4}{3\pi(4l^2-8l+3)}$\\
\hline
-2	&2		&$(1-e^{-2 i \omega })\frac{\sqrt{15} \sqrt[4]{\pi } 2^{l-6}}{\left(\frac{3}{2}-l\right)!} \sqrt{\frac{l! (l-m)!}{\left(2 l^3-3 l^2+l\right) \left(l+\frac{1}{2}\right)! (2 l-4)! (l-m-4)!}}\theta_{l-m-4}\theta_{l-2}$ & $\frac{4}{3\pi(4l^2-8l+3)}$\\
\hline
-1	&all		&$0$ & $\frac{1}{4l(l-1)}$\\
\hline
0	&-2		&$(1-e^{2i\omega})\frac{\sqrt{\frac{15}{2 \pi }} (-1)^l}{8 l (l+1)-6} \sqrt{(l-m+1) (l-m+2) (l+m-1) (l+m)}\theta_{l+m-2}\theta_{l-1}$ & $\frac{4}{\pi(4l^2-1)}$\\
\hline
0	&-1		&$(1-e^{i \omega })\frac{\sqrt{\frac{15}{2 \pi }} (-1)^l}{8 l (l+1)-6} (2 m-1) \sqrt{(l-m+1) (l+m)}\theta_{l-1}\theta_{l+m-1}$ & $\frac{4}{\pi(4l^2-1)}$\\
\hline
0	&0		&$0$ & $\frac{4}{\pi(4l^2-1)}$\\
\hline
0	&1		&$(1-e^{-i\omega})\frac{\sqrt{\frac{15}{2 \pi }} (-1)^{l+1}}{8 l (l+1)-6} (2 m+1) \sqrt{(l-m) (l+m+1)}\theta_{l-1}\theta_{l-m-1}$ & $\frac{4}{\pi(4l^2-1)}$\\
\hline
0	&2		&$(1-e^{-2i\omega})
\frac{\sqrt{\frac{15}{2 \pi }} (-1)^l }{8 l (l+1)-6}\sqrt{(l-m-1) (l-m) (l+m+1) (l+m+2)}
\theta_{l-1}\theta_{l-m-2}$ & $\frac{4}{\pi(4l^2-1)}$\\
\hline
1	&all		&$0$ & $\frac{1}{4l(l+1)}$\\
\hline
2	&-2		&$(1-e^{2i\omega})\frac{\sqrt{15} \sqrt[4]{\pi } 2^{l-4}}{\left(-l-\frac{1}{2}\right)!} \sqrt{\frac{(l+2)! (l-m+4)!}{\left(\left(2 l^3+9 l^2+13 l+6\right) (2 l)! \left(l+\frac{5}{2}\right)!\right) (l-m)!}}
$ & $\frac{4}{3\pi(4l^2+8l+3)}$\\
\hline
2	&-1		&$(1-e^{i\omega})\frac{\sqrt{15} \sqrt[4]{\pi } 2^{l-3}}{\left(-l-\frac{1}{2}\right)!} \sqrt{\frac{l! (l-m+1) (l-m+2) (l-m+3) (l+m+1)}{(2 l+3) (2 l)! \left(l+\frac{5}{2}\right)!}}
$ & $\frac{4}{3\pi(4l^2+8l+3)}$\\
\hline
2	&0		&$0$ &$\frac{4}{3\pi(4l^2+8l+3)}$\\
\hline
2	&1		&$(1-e^{-i\omega})\frac{\sqrt{15} \sqrt[4]{\pi } 2^{l-3}}{\left(-l-\frac{1}{2}\right)!} \sqrt{\frac{l! (l-m+1) (l+m+1) (l+m+2) (l+m+3)}{(2 l+3) (2 l)! \left(\frac{5}{2}-l\right)!}}
$ & $\frac{4}{3\pi(4l^2+8l+3)}$\\
\hline
2	&2		&$(1-e^{-2i\omega})\frac{\sqrt{15} \sqrt[4]{\pi } 2^{l-4}}{\left(-l-\frac{1}{2}\right)!} \sqrt{\frac{l! (l+m+1) (l+m+2) (l+m+3) (l+m+4)}{(2 l+3) (2 l)! \left(\frac{5}{2}-l\right)!}}
$ & $\frac{4}{3\pi(4l^2+8l+3)}$\\
\hline
\hline
\end{tabular}
\end{table*}


\begin{thebibliography}{51}%
\makeatletter
\providecommand \@ifxundefined [1]{%
 \@ifx{#1\undefined}
}%
\providecommand \@ifnum [1]{%
 \ifnum #1\expandafter \@firstoftwo
 \else \expandafter \@secondoftwo
 \fi
}%
\providecommand \@ifx [1]{%
 \ifx #1\expandafter \@firstoftwo
 \else \expandafter \@secondoftwo
 \fi
}%
\providecommand \natexlab [1]{#1}%
\providecommand \enquote  [1]{``#1''}%
\providecommand \bibnamefont  [1]{#1}%
\providecommand \bibfnamefont [1]{#1}%
\providecommand \citenamefont [1]{#1}%
\providecommand \href@noop [0]{\@secondoftwo}%
\providecommand \href [0]{\begingroup \@sanitize@url \@href}%
\providecommand \@href[1]{\@@startlink{#1}\@@href}%
\providecommand \@@href[1]{\endgroup#1\@@endlink}%
\providecommand \@sanitize@url [0]{\catcode `\\12\catcode `\$12\catcode
  `\&12\catcode `\#12\catcode `\^12\catcode `\_12\catcode `\%12\relax}%
\providecommand \@@startlink[1]{}%
\providecommand \@@endlink[0]{}%
\providecommand \url  [0]{\begingroup\@sanitize@url \@url }%
\providecommand \@url [1]{\endgroup\@href {#1}{\urlprefix }}%
\providecommand \urlprefix  [0]{URL }%
\providecommand \Eprint [0]{\href }%
\providecommand \doibase [0]{http://dx.doi.org/}%
\providecommand \selectlanguage [0]{\@gobble}%
\providecommand \bibinfo  [0]{\@secondoftwo}%
\providecommand \bibfield  [0]{\@secondoftwo}%
\providecommand \translation [1]{[#1]}%
\providecommand \BibitemOpen [0]{}%
\providecommand \bibitemStop [0]{}%
\providecommand \bibitemNoStop [0]{.\EOS\space}%
\providecommand \EOS [0]{\spacefactor3000\relax}%
\providecommand \BibitemShut  [1]{\csname bibitem#1\endcsname}%
\let\auto@bib@innerbib\@empty
\bibitem [{\citenamefont {Breuer}\ and\ \citenamefont
  {Petruccione}(2002)}]{Breuer:2002aa}%
  \BibitemOpen
  \bibfield  {author} {\bibinfo {author} {\bibfnamefont {H.~P.}\ \bibnamefont
  {Breuer}}\ and\ \bibinfo {author} {\bibfnamefont {F.}~\bibnamefont
  {Petruccione}},\ }\href@noop {} {\emph {\bibinfo {title} {{The Theory of Open
  Quantum Systems}}}}\ (\bibinfo  {publisher} {Oxford University Press},\
  \bibinfo {address} {Oxford},\ \bibinfo {year} {2002})\BibitemShut {NoStop}%
\bibitem [{\citenamefont {Schlosshauer}(2007)}]{Schlosshauer:2007aa}%
  \BibitemOpen
  \bibfield  {author} {\bibinfo {author} {\bibfnamefont {M.~A.}\ \bibnamefont
  {Schlosshauer}},\ }\href@noop {} {\emph {\bibinfo {title} {{Decoherence and
  the Quantum-To-Classical Transition}}}},\ \bibinfo {edition} {1st}\ ed.\
  (\bibinfo  {publisher} {Springer-Verlag Berlin Heidelberg},\ \bibinfo {year}
  {2007})\BibitemShut {NoStop}%
\bibitem [{\citenamefont {Joos}\ and\ \citenamefont {Zeh}(1985)}]{Joos:1985aa}%
  \BibitemOpen
  \bibfield  {author} {\bibinfo {author} {\bibfnamefont {E.}~\bibnamefont
  {Joos}}\ and\ \bibinfo {author} {\bibfnamefont {H.~D.}\ \bibnamefont {Zeh}},\
  }\href {http://dx.doi.org/10.1007/BF01725541} {\bibfield  {journal} {\bibinfo
   {journal} {Z.~Phys.~B}\ }\textbf {\bibinfo {volume} {59}},\ \bibinfo {pages}
  {223} (\bibinfo {year} {1985})}\BibitemShut {NoStop}%
\bibitem [{\citenamefont {Gallis}\ and\ \citenamefont
  {Fleming}(1990)}]{Gallis:1990aa}%
  \BibitemOpen
  \bibfield  {author} {\bibinfo {author} {\bibfnamefont {M.~R.}\ \bibnamefont
  {Gallis}}\ and\ \bibinfo {author} {\bibfnamefont {G.~N.}\ \bibnamefont
  {Fleming}},\ }\href {http://link.aps.org/doi/10.1103/PhysRevA.42.38}
  {\bibfield  {journal} {\bibinfo  {journal} {Phys.~Rev.~A}\ }\textbf {\bibinfo
  {volume} {42}},\ \bibinfo {pages} {38} (\bibinfo {year} {1990})}\BibitemShut
  {NoStop}%
\bibitem [{\citenamefont {Hornberger}\ and\ \citenamefont
  {Sipe}(2003)}]{Hornberger:2003aa}%
  \BibitemOpen
  \bibfield  {author} {\bibinfo {author} {\bibfnamefont {K.}~\bibnamefont
  {Hornberger}}\ and\ \bibinfo {author} {\bibfnamefont {J.~E.}\ \bibnamefont
  {Sipe}},\ }\href {\doibase 10.1103/PhysRevA.68.012105} {\bibfield  {journal}
  {\bibinfo  {journal} {Phys. Rev. A}\ }\textbf {\bibinfo {volume} {68}},\
  \bibinfo {pages} {012105} (\bibinfo {year} {2003})}\BibitemShut {NoStop}%
\bibitem [{\citenamefont {Hornberger}(2007)}]{Hornberger:2007aa}%
  \BibitemOpen
  \bibfield  {author} {\bibinfo {author} {\bibfnamefont {K.}~\bibnamefont
  {Hornberger}},\ }\href {\doibase 10.1209/0295-5075/77/50007} {\bibfield
  {journal} {\bibinfo  {journal} {Europhysics Letters ({EPL})}\ }\textbf
  {\bibinfo {volume} {77}},\ \bibinfo {pages} {50007} (\bibinfo {year}
  {2007})}\BibitemShut {NoStop}%
\bibitem [{\citenamefont {Gasbarri}\ \emph {et~al.}(2015)\citenamefont
  {Gasbarri}, \citenamefont {Donadi},\ and\ \citenamefont
  {Bassi}}]{Gasbarri:2015aa}%
  \BibitemOpen
  \bibfield  {author} {\bibinfo {author} {\bibfnamefont {G.}~\bibnamefont
  {Gasbarri}}, \bibinfo {author} {\bibfnamefont {S.}~\bibnamefont {Donadi}}, \
  and\ \bibinfo {author} {\bibfnamefont {A.}~\bibnamefont {Bassi}},\ }\href
  {\doibase 10.1088/0143-0807/36/5/055038} {\bibfield  {journal} {\bibinfo
  {journal} {European Journal of Physics}\ }\textbf {\bibinfo {volume} {36}},\
  \bibinfo {pages} {055038} (\bibinfo {year} {2015})}\BibitemShut {NoStop}%
\bibitem [{\citenamefont {Caldeira}\ and\ \citenamefont
  {Leggett}(1983)}]{Caldeira:1983aa}%
  \BibitemOpen
  \bibfield  {author} {\bibinfo {author} {\bibfnamefont {A.~O.}\ \bibnamefont
  {Caldeira}}\ and\ \bibinfo {author} {\bibfnamefont {A.~J.}\ \bibnamefont
  {Leggett}},\ }\href {http://dx.doi.org/10.1016/0378-4371(83)90013-4}
  {\bibfield  {journal} {\bibinfo  {journal} {Phys.~A}\ }\textbf {\bibinfo
  {volume} {121}},\ \bibinfo {pages} {587 } (\bibinfo {year}
  {1983})}\BibitemShut {NoStop}%
\bibitem [{\citenamefont {Hu}\ \emph {et~al.}(1992)\citenamefont {Hu},
  \citenamefont {Paz},\ and\ \citenamefont {Zhang}}]{Hu:1992aa}%
  \BibitemOpen
  \bibfield  {author} {\bibinfo {author} {\bibfnamefont {B.~L.}\ \bibnamefont
  {Hu}}, \bibinfo {author} {\bibfnamefont {J.~P.}\ \bibnamefont {Paz}}, \ and\
  \bibinfo {author} {\bibfnamefont {Y.}~\bibnamefont {Zhang}},\ }\href
  {\doibase 10.1103/PhysRevD.45.2843} {\bibfield  {journal} {\bibinfo
  {journal} {Phys.~Rev.~D}\ }\textbf {\bibinfo {volume} {45}},\ \bibinfo
  {pages} {2843} (\bibinfo {year} {1992})}\BibitemShut {NoStop}%
\bibitem [{\citenamefont {Ferialdi}(2017)}]{Ferialdi:2017aa}%
  \BibitemOpen
  \bibfield  {author} {\bibinfo {author} {\bibfnamefont {L.}~\bibnamefont
  {Ferialdi}},\ }\href {\doibase 10.1103/PhysRevA.95.052109} {\bibfield
  {journal} {\bibinfo  {journal} {Phys. Rev. A}\ }\textbf {\bibinfo {volume}
  {95}},\ \bibinfo {pages} {052109} (\bibinfo {year} {2017})}\BibitemShut
  {NoStop}%
\bibitem [{\citenamefont {Carlesso}\ and\ \citenamefont
  {Bassi}(2017)}]{Carlesso:2017aa}%
  \BibitemOpen
  \bibfield  {author} {\bibinfo {author} {\bibfnamefont {M.}~\bibnamefont
  {Carlesso}}\ and\ \bibinfo {author} {\bibfnamefont {A.}~\bibnamefont
  {Bassi}},\ }\href {\doibase 10.1103/PhysRevA.95.052119} {\bibfield  {journal}
  {\bibinfo  {journal} {Phys.~Rev.~A}\ }\textbf {\bibinfo {volume} {95}},\
  \bibinfo {pages} {052119} (\bibinfo {year} {2017})}\BibitemShut {NoStop}%
\bibitem [{\citenamefont {Hornberger}\ \emph {et~al.}(2004)\citenamefont
  {Hornberger}, \citenamefont {Sipe},\ and\ \citenamefont
  {Arndt}}]{Hornberger:2004aa}%
  \BibitemOpen
  \bibfield  {author} {\bibinfo {author} {\bibfnamefont {K.}~\bibnamefont
  {Hornberger}}, \bibinfo {author} {\bibfnamefont {J.~E.}\ \bibnamefont
  {Sipe}}, \ and\ \bibinfo {author} {\bibfnamefont {M.}~\bibnamefont {Arndt}},\
  }\href {\doibase 10.1103/PhysRevA.70.053608} {\bibfield  {journal} {\bibinfo
  {journal} {Phys.~Rev.~A}\ }\textbf {\bibinfo {volume} {70}},\ \bibinfo
  {pages} {053608} (\bibinfo {year} {2004})}\BibitemShut {NoStop}%
\bibitem [{\citenamefont {Bateman}\ \emph {et~al.}(2014)\citenamefont
  {Bateman}, \citenamefont {Nimmrichter}, \citenamefont {Hornberger},\ and\
  \citenamefont {Ulbricht}}]{Bateman:2014aa}%
  \BibitemOpen
  \bibfield  {author} {\bibinfo {author} {\bibfnamefont {J.}~\bibnamefont
  {Bateman}}, \bibinfo {author} {\bibfnamefont {S.}~\bibnamefont
  {Nimmrichter}}, \bibinfo {author} {\bibfnamefont {K.}~\bibnamefont
  {Hornberger}}, \ and\ \bibinfo {author} {\bibfnamefont {H.}~\bibnamefont
  {Ulbricht}},\ }\href {http://dx.doi.org/10.1038/ncomms5788} {\bibfield
  {journal} {\bibinfo  {journal} {Nat.~Commun.}\ }\textbf {\bibinfo {volume}
  {5}} (\bibinfo {year} {2014})}\BibitemShut {NoStop}%
\bibitem [{\citenamefont {Belenchia}\ \emph {et~al.}(2019)\citenamefont
  {Belenchia}, \citenamefont {Gasbarri}, \citenamefont {Kaltenbaek},
  \citenamefont {Ulbricht},\ and\ \citenamefont
  {Paternostro}}]{Belenchia:2019aa}%
  \BibitemOpen
  \bibfield  {author} {\bibinfo {author} {\bibfnamefont {A.}~\bibnamefont
  {Belenchia}}, \bibinfo {author} {\bibfnamefont {G.}~\bibnamefont {Gasbarri}},
  \bibinfo {author} {\bibfnamefont {R.}~\bibnamefont {Kaltenbaek}}, \bibinfo
  {author} {\bibfnamefont {H.}~\bibnamefont {Ulbricht}}, \ and\ \bibinfo
  {author} {\bibfnamefont {M.}~\bibnamefont {Paternostro}},\ }\href {\doibase
  10.1103/PhysRevA.100.033813} {\bibfield  {journal} {\bibinfo  {journal}
  {Phys. Rev. A}\ }\textbf {\bibinfo {volume} {100}},\ \bibinfo {pages}
  {033813} (\bibinfo {year} {2019})}\BibitemShut {NoStop}%
\bibitem [{\citenamefont {Kovachy}\ \emph {et~al.}(2015)\citenamefont {Kovachy}
  \emph {et~al.}}]{Kovachy:2015aa}%
  \BibitemOpen
  \bibfield  {author} {\bibinfo {author} {\bibfnamefont {T.}~\bibnamefont
  {Kovachy}} \emph {et~al.},\ }\href {http://dx.doi.org/10.1038/nature16155}
  {\bibfield  {journal} {\bibinfo  {journal} {Nature}\ }\textbf {\bibinfo
  {volume} {528}},\ \bibinfo {pages} {530} (\bibinfo {year}
  {2015})}\BibitemShut {NoStop}%
\bibitem [{\citenamefont {Becker}\ \emph {et~al.}(2018)\citenamefont {Becker}
  \emph {et~al.}}]{Becker:2018aa}%
  \BibitemOpen
  \bibfield  {author} {\bibinfo {author} {\bibfnamefont {D.}~\bibnamefont
  {Becker}} \emph {et~al.},\ }\href {\doibase 10.1038/s41586-018-0605-1}
  {\bibfield  {journal} {\bibinfo  {journal} {Nature}\ }\textbf {\bibinfo
  {volume} {562}},\ \bibinfo {pages} {391} (\bibinfo {year}
  {2018})}\BibitemShut {NoStop}%
\bibitem [{\citenamefont {Aspelmeyer}\ \emph {et~al.}(2014)\citenamefont
  {Aspelmeyer}, \citenamefont {Kippenberg},\ and\ \citenamefont
  {Marquardt}}]{Aspelmeyer:2014aa}%
  \BibitemOpen
  \bibfield  {author} {\bibinfo {author} {\bibfnamefont {M.}~\bibnamefont
  {Aspelmeyer}}, \bibinfo {author} {\bibfnamefont {T.~J.}\ \bibnamefont
  {Kippenberg}}, \ and\ \bibinfo {author} {\bibfnamefont {F.}~\bibnamefont
  {Marquardt}},\ }\href {https://link.aps.org/doi/10.1103/RevModPhys.86.1391}
  {\bibfield  {journal} {\bibinfo  {journal} {Rev.~Mod.~Phys.}\ }\textbf
  {\bibinfo {volume} {86}},\ \bibinfo {pages} {1391} (\bibinfo {year}
  {2014})}\BibitemShut {NoStop}%
\bibitem [{\citenamefont {Abbott}\ \emph {et~al.}(2016)\citenamefont {Abbott}
  \emph {et~al.}}]{Abbott:2016aa}%
  \BibitemOpen
  \bibfield  {author} {\bibinfo {author} {\bibfnamefont {B.~P.}\ \bibnamefont
  {Abbott}} \emph {et~al.} (\bibinfo {collaboration} {LIGO Scientific
  Collaboration and Virgo Collaboration}),\ }\href
  {http://link.aps.org/doi/10.1103/PhysRevLett.116.061102} {\bibfield
  {journal} {\bibinfo  {journal} {Phys.~Rev.~Lett.}\ }\textbf {\bibinfo
  {volume} {116}},\ \bibinfo {pages} {061102} (\bibinfo {year}
  {2016})}\BibitemShut {NoStop}%
\bibitem [{\citenamefont {Armano}\ \emph {et~al.}(2016)\citenamefont {Armano}
  \emph {et~al.}}]{Armano:2016aa}%
  \BibitemOpen
  \bibfield  {author} {\bibinfo {author} {\bibfnamefont {M.}~\bibnamefont
  {Armano}} \emph {et~al.},\ }\href
  {http://link.aps.org/doi/10.1103/PhysRevLett.116.231101} {\bibfield
  {journal} {\bibinfo  {journal} {Phys.~Rev.~Lett.}\ }\textbf {\bibinfo
  {volume} {116}},\ \bibinfo {pages} {231101} (\bibinfo {year}
  {2016})}\BibitemShut {NoStop}%
\bibitem [{\citenamefont {Vovrosh}\ \emph {et~al.}(2017)\citenamefont {Vovrosh}
  \emph {et~al.}}]{Vovrosh:2017aa}%
  \BibitemOpen
  \bibfield  {author} {\bibinfo {author} {\bibfnamefont {J.}~\bibnamefont
  {Vovrosh}} \emph {et~al.},\ }\href
  {http://josab.osa.org/abstract.cfm?URI=josab-34-7-1421} {\bibfield  {journal}
  {\bibinfo  {journal} {J.~Opt.~Soc.~Am.~B}\ }\textbf {\bibinfo {volume}
  {34}},\ \bibinfo {pages} {1421} (\bibinfo {year} {2017})}\BibitemShut
  {NoStop}%
\bibitem [{\citenamefont {Vinante}\ \emph {et~al.}(2017)\citenamefont
  {Vinante}, \citenamefont {Mezzena}, \citenamefont {Falferi}, \citenamefont
  {Carlesso},\ and\ \citenamefont {Bassi}}]{Vinante:2017aa}%
  \BibitemOpen
  \bibfield  {author} {\bibinfo {author} {\bibfnamefont {A.}~\bibnamefont
  {Vinante}}, \bibinfo {author} {\bibfnamefont {R.}~\bibnamefont {Mezzena}},
  \bibinfo {author} {\bibfnamefont {P.}~\bibnamefont {Falferi}}, \bibinfo
  {author} {\bibfnamefont {M.}~\bibnamefont {Carlesso}}, \ and\ \bibinfo
  {author} {\bibfnamefont {A.}~\bibnamefont {Bassi}},\ }\href
  {https://link.aps.org/doi/10.1103/PhysRevLett.119.110401} {\bibfield
  {journal} {\bibinfo  {journal} {Phys.~Rev.~Lett.}\ }\textbf {\bibinfo
  {volume} {119}},\ \bibinfo {pages} {110401} (\bibinfo {year}
  {2017})}\BibitemShut {NoStop}%
\bibitem [{\citenamefont {Hempston}\ \emph {et~al.}(2017)\citenamefont
  {Hempston}, \citenamefont {Vovrosh}, \citenamefont {Toro{\v s}},
  \citenamefont {Winstone}, \citenamefont {Rashid},\ and\ \citenamefont
  {Ulbricht}}]{Hempston:2017aa}%
  \BibitemOpen
  \bibfield  {author} {\bibinfo {author} {\bibfnamefont {D.}~\bibnamefont
  {Hempston}}, \bibinfo {author} {\bibfnamefont {J.}~\bibnamefont {Vovrosh}},
  \bibinfo {author} {\bibfnamefont {M.}~\bibnamefont {Toro{\v s}}}, \bibinfo
  {author} {\bibfnamefont {G.}~\bibnamefont {Winstone}}, \bibinfo {author}
  {\bibfnamefont {M.}~\bibnamefont {Rashid}}, \ and\ \bibinfo {author}
  {\bibfnamefont {H.}~\bibnamefont {Ulbricht}},\ }\href
  {https://doi.org/10.1063/1.4993555} {\bibfield  {journal} {\bibinfo
  {journal} {Applied Physics Letters}\ }\textbf {\bibinfo {volume} {111}},\
  \bibinfo {pages} {133111} (\bibinfo {year} {2017})}\BibitemShut {NoStop}%
\bibitem [{\citenamefont {Vinante}\ \emph {et~al.}(2019)\citenamefont
  {Vinante}, \citenamefont {Pontin}, \citenamefont {Rashid}, \citenamefont
  {Toro\ifmmode~\check{s}\else \v{s}\fi{}}, \citenamefont {Barker},\ and\
  \citenamefont {Ulbricht}}]{Vinante:2019aa}%
  \BibitemOpen
  \bibfield  {author} {\bibinfo {author} {\bibfnamefont {A.}~\bibnamefont
  {Vinante}}, \bibinfo {author} {\bibfnamefont {A.}~\bibnamefont {Pontin}},
  \bibinfo {author} {\bibfnamefont {M.}~\bibnamefont {Rashid}}, \bibinfo
  {author} {\bibfnamefont {M.}~\bibnamefont {Toro\ifmmode~\check{s}\else
  \v{s}\fi{}}}, \bibinfo {author} {\bibfnamefont {P.~F.}\ \bibnamefont
  {Barker}}, \ and\ \bibinfo {author} {\bibfnamefont {H.}~\bibnamefont
  {Ulbricht}},\ }\href {\doibase 10.1103/PhysRevA.100.012119} {\bibfield
  {journal} {\bibinfo  {journal} {Phys. Rev. A}\ }\textbf {\bibinfo {volume}
  {100}},\ \bibinfo {pages} {012119} (\bibinfo {year} {2019})}\BibitemShut
  {NoStop}%
  \bibitem [{\citenamefont {Gardiner}\ and\ \citenamefont
  {Zoller}(2004)}]{Gardiner:2004aa}%
  \BibitemOpen
  \bibfield  {author} {\bibinfo {author} {\bibfnamefont {C.}~\bibnamefont
  {Gardiner}}\ and\ \bibinfo {author} {\bibfnamefont {P.}~\bibnamefont
  {Zoller}},\ }\href@noop {} {\emph {\bibinfo {title} {{Quantum Noise}}}}\
  (\bibinfo  {publisher} {Springer-Verlag Berlin Heidelberg},\ \bibinfo {year}
  {2004})\BibitemShut {NoStop}%
\bibitem [{\citenamefont {Clerk}\ \emph {et~al.}(2010)\citenamefont {Clerk},
  \citenamefont {Devoret}, \citenamefont {Girvin}, \citenamefont {Marquardt},\
  and\ \citenamefont {Schoelkopf}}]{Clerk:2010aa}%
  \BibitemOpen
  \bibfield  {author} {\bibinfo {author} {\bibfnamefont {A.~A.}\ \bibnamefont
  {Clerk}}, \bibinfo {author} {\bibfnamefont {M.~H.}\ \bibnamefont {Devoret}},
  \bibinfo {author} {\bibfnamefont {S.~M.}\ \bibnamefont {Girvin}}, \bibinfo
  {author} {\bibfnamefont {F.}~\bibnamefont {Marquardt}}, \ and\ \bibinfo
  {author} {\bibfnamefont {R.~J.}\ \bibnamefont {Schoelkopf}},\ }\href
  {\doibase 10.1103/RevModPhys.82.1155} {\bibfield  {journal} {\bibinfo
  {journal} {Rev. Mod. Phys.}\ }\textbf {\bibinfo {volume} {82}},\ \bibinfo
  {pages} {1155} (\bibinfo {year} {2010})}\BibitemShut {NoStop}%
\bibitem [{\citenamefont {Paterson}\ \emph {et~al.}(2001)\citenamefont
  {Paterson}, \citenamefont {MacDonald}, \citenamefont {Arlt}, \citenamefont
  {Sibbett}, \citenamefont {Bryant},\ and\ \citenamefont
  {Dholakia}}]{Paterson:2001aa}%
  \BibitemOpen
  \bibfield  {author} {\bibinfo {author} {\bibfnamefont {L.}~\bibnamefont
  {Paterson}}, \bibinfo {author} {\bibfnamefont {M.~P.}\ \bibnamefont
  {MacDonald}}, \bibinfo {author} {\bibfnamefont {J.}~\bibnamefont {Arlt}},
  \bibinfo {author} {\bibfnamefont {W.}~\bibnamefont {Sibbett}}, \bibinfo
  {author} {\bibfnamefont {P.~E.}\ \bibnamefont {Bryant}}, \ and\ \bibinfo
  {author} {\bibfnamefont {K.}~\bibnamefont {Dholakia}},\ }\href {\doibase
  10.1126/science.1058591} {\bibfield  {journal} {\bibinfo  {journal}
  {Science}\ }\textbf {\bibinfo {volume} {292}},\ \bibinfo {pages} {912}
  (\bibinfo {year} {2001})}\BibitemShut {NoStop}%
\bibitem [{\citenamefont {Bonin}\ \emph {et~al.}(2002)\citenamefont {Bonin},
  \citenamefont {Kourmanov},\ and\ \citenamefont {Walker}}]{Bonin:2002aa}%
  \BibitemOpen
  \bibfield  {author} {\bibinfo {author} {\bibfnamefont {K.~D.}\ \bibnamefont
  {Bonin}}, \bibinfo {author} {\bibfnamefont {B.}~\bibnamefont {Kourmanov}}, \
  and\ \bibinfo {author} {\bibfnamefont {T.~G.}\ \bibnamefont {Walker}},\
  }\href {http://www.opticsexpress.org/abstract.cfm?URI=oe-10-19-984}
  {\bibfield  {journal} {\bibinfo  {journal} {Opt. Express}\ }\textbf {\bibinfo
  {volume} {10}},\ \bibinfo {pages} {984} (\bibinfo {year} {2002})}\BibitemShut
  {NoStop}%
\bibitem [{\citenamefont {Shelton}\ \emph {et~al.}(2005)\citenamefont
  {Shelton}, \citenamefont {Bonin},\ and\ \citenamefont
  {Walker}}]{Shelton:2005aa}%
  \BibitemOpen
  \bibfield  {author} {\bibinfo {author} {\bibfnamefont {W.~A.}\ \bibnamefont
  {Shelton}}, \bibinfo {author} {\bibfnamefont {K.~D.}\ \bibnamefont {Bonin}},
  \ and\ \bibinfo {author} {\bibfnamefont {T.~G.}\ \bibnamefont {Walker}},\
  }\href {https://link.aps.org/doi/10.1103/PhysRevE.71.036204} {\bibfield
  {journal} {\bibinfo  {journal} {Phys.~Rev.~E}\ }\textbf {\bibinfo {volume}
  {71}},\ \bibinfo {pages} {036204} (\bibinfo {year} {2005})}\BibitemShut
  {NoStop}%
\bibitem [{\citenamefont {Jones}\ \emph {et~al.}(2009)\citenamefont {Jones}
  \emph {et~al.}}]{Jones:2009aa}%
  \BibitemOpen
  \bibfield  {author} {\bibinfo {author} {\bibfnamefont {P.~H.}\ \bibnamefont
  {Jones}} \emph {et~al.},\ }\href {http://dx.doi.org/10.1021/nn900818n}
  {\bibfield  {journal} {\bibinfo  {journal} {ACS Nano}\ }\textbf {\bibinfo
  {volume} {3}},\ \bibinfo {pages} {3077} (\bibinfo {year} {2009})}\BibitemShut
  {NoStop}%
\bibitem [{\citenamefont {Tong}\ \emph {et~al.}(2010)\citenamefont {Tong},
  \citenamefont {Miljkovi{\'c}},\ and\ \citenamefont {K{\"a}ll}}]{Tong:2010aa}%
  \BibitemOpen
  \bibfield  {author} {\bibinfo {author} {\bibfnamefont {L.}~\bibnamefont
  {Tong}}, \bibinfo {author} {\bibfnamefont {V.~D.}\ \bibnamefont
  {Miljkovi{\'c}}}, \ and\ \bibinfo {author} {\bibfnamefont {M.}~\bibnamefont
  {K{\"a}ll}},\ }\href {http://dx.doi.org/10.1021/nl9034434} {\bibfield
  {journal} {\bibinfo  {journal} {Nano Lett.}\ }\textbf {\bibinfo {volume}
  {10}},\ \bibinfo {pages} {268} (\bibinfo {year} {2010})}\BibitemShut
  {NoStop}%
\bibitem [{\citenamefont {Arita}\ \emph {et~al.}(2013)\citenamefont {Arita},
  \citenamefont {Mazilu},\ and\ \citenamefont {Dholakia}}]{Arita:2013aa}%
  \BibitemOpen
  \bibfield  {author} {\bibinfo {author} {\bibfnamefont {Y.}~\bibnamefont
  {Arita}}, \bibinfo {author} {\bibfnamefont {M.}~\bibnamefont {Mazilu}}, \
  and\ \bibinfo {author} {\bibfnamefont {K.}~\bibnamefont {Dholakia}},\ }\href
  {http://dx.doi.org/10.1038/ncomms3374} {\bibfield  {journal} {\bibinfo
  {journal} {Nat.~Commun.}\ }\textbf {\bibinfo {volume} {4}},\ \bibinfo {pages}
  {2374} (\bibinfo {year} {2013})}\BibitemShut {NoStop}%
\bibitem [{\citenamefont {Kuhn}\ \emph {et~al.}(2015)\citenamefont {Kuhn} \emph
  {et~al.}}]{Kuhn:2015aa}%
  \BibitemOpen
  \bibfield  {author} {\bibinfo {author} {\bibfnamefont {S.}~\bibnamefont
  {Kuhn}} \emph {et~al.},\ }\href
  {http://dx.doi.org/10.1021/acs.nanolett.5b02302} {\bibfield  {journal}
  {\bibinfo  {journal} {Nano Lett.}\ }\textbf {\bibinfo {volume} {15}},\
  \bibinfo {pages} {5604} (\bibinfo {year} {2015})}\BibitemShut {NoStop}%
\bibitem [{\citenamefont {Hoang}\ \emph {et~al.}(2016)\citenamefont {Hoang}
  \emph {et~al.}}]{Hoang:2016ab}%
  \BibitemOpen
  \bibfield  {author} {\bibinfo {author} {\bibfnamefont {T.~M.}\ \bibnamefont
  {Hoang}} \emph {et~al.},\ }\href
  {https://link.aps.org/doi/10.1103/PhysRevLett.117.123604} {\bibfield
  {journal} {\bibinfo  {journal} {Phys.~Rev.~Lett.}\ }\textbf {\bibinfo
  {volume} {117}},\ \bibinfo {pages} {123604} (\bibinfo {year}
  {2016})}\BibitemShut {NoStop}%
\bibitem [{\citenamefont {Kuhn}\ \emph {et~al.}(2017)\citenamefont {Kuhn} \emph
  {et~al.}}]{Kuhn:2017ab}%
  \BibitemOpen
  \bibfield  {author} {\bibinfo {author} {\bibfnamefont {S.}~\bibnamefont
  {Kuhn}} \emph {et~al.},\ }\href
  {http://www.osapublishing.org/optica/abstract.cfm?URI=optica-4-3-356}
  {\bibfield  {journal} {\bibinfo  {journal} {Optica}\ }\textbf {\bibinfo
  {volume} {4}},\ \bibinfo {pages} {356} (\bibinfo {year} {2017})}\BibitemShut
  {NoStop}%
\bibitem [{\citenamefont {Rashid}\ \emph {et~al.}(2018)\citenamefont {Rashid},
  \citenamefont {Toro\ifmmode~\check{s}\else \v{s}\fi{}}, \citenamefont
  {Setter},\ and\ \citenamefont {Ulbricht}}]{Rashid:2018aa}%
  \BibitemOpen
  \bibfield  {author} {\bibinfo {author} {\bibfnamefont {M.}~\bibnamefont
  {Rashid}}, \bibinfo {author} {\bibfnamefont {M.}~\bibnamefont
  {Toro\ifmmode~\check{s}\else \v{s}\fi{}}}, \bibinfo {author} {\bibfnamefont
  {A.}~\bibnamefont {Setter}}, \ and\ \bibinfo {author} {\bibfnamefont
  {H.}~\bibnamefont {Ulbricht}},\ }\href {\doibase
  10.1103/PhysRevLett.121.253601} {\bibfield  {journal} {\bibinfo  {journal}
  {Phys. Rev. Lett.}\ }\textbf {\bibinfo {volume} {121}},\ \bibinfo {pages}
  {253601} (\bibinfo {year} {2018})}\BibitemShut {NoStop}%
\bibitem [{\citenamefont {Carlesso}\ \emph {et~al.}(2017)\citenamefont
  {Carlesso}, \citenamefont {Paternostro}, \citenamefont {Ulbricht},\ and\
  \citenamefont {Bassi}}]{Carlesso:2017ac}%
  \BibitemOpen
  \bibfield  {author} {\bibinfo {author} {\bibfnamefont {M.}~\bibnamefont
  {Carlesso}}, \bibinfo {author} {\bibfnamefont {M.}~\bibnamefont
  {Paternostro}}, \bibinfo {author} {\bibfnamefont {H.}~\bibnamefont
  {Ulbricht}}, \ and\ \bibinfo {author} {\bibfnamefont {A.}~\bibnamefont
  {Bassi}},\ }\href {https://arxiv.org/abs/1710.08695} {\bibfield  {journal}
  {\bibinfo  {journal} {ArXiv}\ } (\bibinfo {year} {2017})},\ \Eprint
  {http://arxiv.org/abs/1710.08695} {1710.08695} \BibitemShut {NoStop}%
\bibitem [{\citenamefont {Schrinski}\ \emph {et~al.}(2017)\citenamefont
  {Schrinski}, \citenamefont {Stickler},\ and\ \citenamefont
  {Hornberger}}]{Schrinski:2017aa}%
  \BibitemOpen
  \bibfield  {author} {\bibinfo {author} {\bibfnamefont {B.}~\bibnamefont
  {Schrinski}}, \bibinfo {author} {\bibfnamefont {B.~A.}\ \bibnamefont
  {Stickler}}, \ and\ \bibinfo {author} {\bibfnamefont {K.}~\bibnamefont
  {Hornberger}},\ }\href {http://josab.osa.org/abstract.cfm?URI=josab-34-6-C1}
  {\bibfield  {journal} {\bibinfo  {journal} {J.~Opt.~Soc.~Am.~B}\ }\textbf
  {\bibinfo {volume} {34}},\ \bibinfo {pages} {C1} (\bibinfo {year}
  {2017})}\BibitemShut {NoStop}%
\bibitem [{\citenamefont {Carlesso}\ \emph {et~al.}(2018)\citenamefont
  {Carlesso}, \citenamefont {Paternostro}, \citenamefont {Ulbricht},
  \citenamefont {Vinante},\ and\ \citenamefont {Bassi}}]{Carlesso:2018ab}%
  \BibitemOpen
  \bibfield  {author} {\bibinfo {author} {\bibfnamefont {M.}~\bibnamefont
  {Carlesso}}, \bibinfo {author} {\bibfnamefont {M.}~\bibnamefont
  {Paternostro}}, \bibinfo {author} {\bibfnamefont {H.}~\bibnamefont
  {Ulbricht}}, \bibinfo {author} {\bibfnamefont {A.}~\bibnamefont {Vinante}}, \
  and\ \bibinfo {author} {\bibfnamefont {A.}~\bibnamefont {Bassi}},\ }\href
  {http://iopscience.iop.org/article/10.1088/1367-2630/aad863/meta} {\bibfield
  {journal} {\bibinfo  {journal} {New J.~Phys.}\ }\textbf {\bibinfo {volume}
  {20}},\ \bibinfo {pages} {083022} (\bibinfo {year} {2018})}\BibitemShut
  {NoStop}%
\bibitem [{\citenamefont {Stickler}\ \emph {et~al.}(2018)\citenamefont
  {Stickler}, \citenamefont {Schrinski},\ and\ \citenamefont
  {Hornberger}}]{Stickler:2018aa}%
  \BibitemOpen
  \bibfield  {author} {\bibinfo {author} {\bibfnamefont {B.~A.}\ \bibnamefont
  {Stickler}}, \bibinfo {author} {\bibfnamefont {B.}~\bibnamefont {Schrinski}},
  \ and\ \bibinfo {author} {\bibfnamefont {K.}~\bibnamefont {Hornberger}},\
  }\href {\doibase 10.1103/PhysRevLett.121.040401} {\bibfield  {journal}
  {\bibinfo  {journal} {Phys. Rev. Lett.}\ }\textbf {\bibinfo {volume} {121}},\
  \bibinfo {pages} {040401} (\bibinfo {year} {2018})}\BibitemShut {NoStop}%
\bibitem [{\citenamefont {Bhattacharya}\ and\ \citenamefont
  {Meystre}(2007)}]{Bhattacharya:2007aa}%
  \BibitemOpen
  \bibfield  {author} {\bibinfo {author} {\bibfnamefont {M.}~\bibnamefont
  {Bhattacharya}}\ and\ \bibinfo {author} {\bibfnamefont {P.}~\bibnamefont
  {Meystre}},\ }\href {http://link.aps.org/doi/10.1103/PhysRevLett.99.153603}
  {\bibfield  {journal} {\bibinfo  {journal} {Phys.~Rev.~Lett.}\ }\textbf
  {\bibinfo {volume} {99}},\ \bibinfo {pages} {153603} (\bibinfo {year}
  {2007})}\BibitemShut {NoStop}%
\bibitem [{\citenamefont {Trojek}\ \emph {et~al.}(2012)\citenamefont {Trojek},
  \citenamefont {Chv\'{a}tal},\ and\ \citenamefont
  {Zem\'{a}nek}}]{Trojek:2012aa}%
  \BibitemOpen
  \bibfield  {author} {\bibinfo {author} {\bibfnamefont {J.}~\bibnamefont
  {Trojek}}, \bibinfo {author} {\bibfnamefont {L.}~\bibnamefont {Chv\'{a}tal}},
  \ and\ \bibinfo {author} {\bibfnamefont {P.}~\bibnamefont {Zem\'{a}nek}},\
  }\href {http://josaa.osa.org/abstract.cfm?URI=josaa-29-7-1224} {\bibfield
  {journal} {\bibinfo  {journal} {J.~Opt.~Soc.~Am.~A}\ }\textbf {\bibinfo
  {volume} {29}},\ \bibinfo {pages} {1224} (\bibinfo {year}
  {2012})}\BibitemShut {NoStop}%
\bibitem [{\citenamefont {Bhattacharya}(2015)}]{Bhattacharya:2015aa}%
  \BibitemOpen
  \bibfield  {author} {\bibinfo {author} {\bibfnamefont {M.}~\bibnamefont
  {Bhattacharya}},\ }\href {\doibase 10.1364/JOSAB.32.000B55} {\bibfield
  {journal} {\bibinfo  {journal} {J. Opt. Soc. Am. B}\ }\textbf {\bibinfo
  {volume} {32}},\ \bibinfo {pages} {B55} (\bibinfo {year} {2015})}\BibitemShut
  {NoStop}%
\bibitem [{\citenamefont {Shi}\ and\ \citenamefont
  {Bhattacharya}(2016)}]{Shi:2016aa}%
  \BibitemOpen
  \bibfield  {author} {\bibinfo {author} {\bibfnamefont {H.}~\bibnamefont
  {Shi}}\ and\ \bibinfo {author} {\bibfnamefont {M.}~\bibnamefont
  {Bhattacharya}},\ }\href {http://stacks.iop.org/0953-4075/49/i=15/a=153001}
  {\bibfield  {journal} {\bibinfo  {journal} {J.~Phys.~B}\ }\textbf {\bibinfo
  {volume} {49}},\ \bibinfo {pages} {153001} (\bibinfo {year}
  {2016})}\BibitemShut {NoStop}%
\bibitem [{\citenamefont {Stickler}\ \emph
  {et~al.}(2016{\natexlab{a}})\citenamefont {Stickler} \emph
  {et~al.}}]{Stickler:2016ab}%
  \BibitemOpen
  \bibfield  {author} {\bibinfo {author} {\bibfnamefont {B.~A.}\ \bibnamefont
  {Stickler}} \emph {et~al.},\ }\href
  {https://link.aps.org/doi/10.1103/PhysRevA.94.033818} {\bibfield  {journal}
  {\bibinfo  {journal} {Phys.~Rev.~A}\ }\textbf {\bibinfo {volume} {94}},\
  \bibinfo {pages} {033818} (\bibinfo {year} {2016}{\natexlab{a}})}\BibitemShut
  {NoStop}%
\bibitem [{\citenamefont {Zhong}\ and\ \citenamefont
  {Robicheaux}(2016)}]{Zhong:2016aa}%
  \BibitemOpen
  \bibfield  {author} {\bibinfo {author} {\bibfnamefont {C.}~\bibnamefont
  {Zhong}}\ and\ \bibinfo {author} {\bibfnamefont {F.}~\bibnamefont
  {Robicheaux}},\ }\href {http://link.aps.org/doi/10.1103/PhysRevA.94.052109}
  {\bibfield  {journal} {\bibinfo  {journal} {Phys.~Rev.~A}\ }\textbf {\bibinfo
  {volume} {94}},\ \bibinfo {pages} {052109} (\bibinfo {year}
  {2016})}\BibitemShut {NoStop}%
\bibitem [{\citenamefont {Stickler}\ \emph
  {et~al.}(2016{\natexlab{b}})\citenamefont {Stickler}, \citenamefont
  {Papendell},\ and\ \citenamefont {Hornberger}}]{Stickler:2016aa}%
  \BibitemOpen
  \bibfield  {author} {\bibinfo {author} {\bibfnamefont {B.~A.}\ \bibnamefont
  {Stickler}}, \bibinfo {author} {\bibfnamefont {B.}~\bibnamefont {Papendell}},
  \ and\ \bibinfo {author} {\bibfnamefont {K.}~\bibnamefont {Hornberger}},\
  }\href {http://link.aps.org/doi/10.1103/PhysRevA.94.033828} {\bibfield
  {journal} {\bibinfo  {journal} {Phys.~Rev.~A}\ }\textbf {\bibinfo {volume}
  {94}},\ \bibinfo {pages} {033828} (\bibinfo {year}
  {2016}{\natexlab{b}})}\BibitemShut {NoStop}%
\bibitem [{\citenamefont {Papendell}\ \emph {et~al.}(2017)\citenamefont
  {Papendell}, \citenamefont {Stickler},\ and\ \citenamefont
  {Hornberger}}]{Papendell:2017aa}%
  \BibitemOpen
  \bibfield  {author} {\bibinfo {author} {\bibfnamefont {B.}~\bibnamefont
  {Papendell}}, \bibinfo {author} {\bibfnamefont {B.~A.}\ \bibnamefont
  {Stickler}}, \ and\ \bibinfo {author} {\bibfnamefont {K.}~\bibnamefont
  {Hornberger}},\ }\href {\doibase 10.1088/1367-2630/aa99d1} {\bibfield
  {journal} {\bibinfo  {journal} {New Journal of Physics}\ }\textbf {\bibinfo
  {volume} {19}},\ \bibinfo {pages} {122001} (\bibinfo {year}
  {2017})}\BibitemShut {NoStop}%
\bibitem [{\citenamefont {Fischer}\ \emph {et~al.}(2013)\citenamefont
  {Fischer}, \citenamefont {Gneiting},\ and\ \citenamefont
  {Hornberger}}]{Fischer:2013aa}%
  \BibitemOpen
  \bibfield  {author} {\bibinfo {author} {\bibfnamefont {T.}~\bibnamefont
  {Fischer}}, \bibinfo {author} {\bibfnamefont {C.}~\bibnamefont {Gneiting}}, \
  and\ \bibinfo {author} {\bibfnamefont {K.}~\bibnamefont {Hornberger}},\
  }\href {http://stacks.iop.org/1367-2630/15/i=6/a=063004} {\bibfield
  {journal} {\bibinfo  {journal} {New J.~Phys.}\ }\textbf {\bibinfo {volume}
  {15}},\ \bibinfo {pages} {063004} (\bibinfo {year} {2013})}\BibitemShut
  {NoStop}%
\bibitem [{\citenamefont {Sakurai}\ and\ \citenamefont
  {Napolitano}(2011)}]{Sakurai:2011aa}%
  \BibitemOpen
  \bibfield  {author} {\bibinfo {author} {\bibfnamefont {J.}~\bibnamefont
  {Sakurai}}\ and\ \bibinfo {author} {\bibfnamefont {J.}~\bibnamefont
  {Napolitano}},\ }\href@noop {} {\emph {\bibinfo {title} {Modern Quantum
  Mechanics}}}\ (\bibinfo  {publisher} {Addison-Wesley},\ \bibinfo {year}
  {2011})\BibitemShut {NoStop}%
\bibitem [{\citenamefont {Kranendonk}(1963)}]{Kranendonk:1963aa}%
  \BibitemOpen
  \bibfield  {author} {\bibinfo {author} {\bibfnamefont {J.~V.}\ \bibnamefont
  {Kranendonk}},\ }\href {\doibase 10.1139/p63-047} {\bibfield  {journal}
  {\bibinfo  {journal} {Can. J. Phys.}\ }\textbf {\bibinfo {volume} {41}},\
  \bibinfo {pages} {433} (\bibinfo {year} {1963})}\BibitemShut {NoStop}%
\bibitem [{\citenamefont {Zettili}(2009)}]{Zettili:2009aa}%
  \BibitemOpen
  \bibfield  {author} {\bibinfo {author} {\bibfnamefont {N.}~\bibnamefont
  {Zettili}},\ }\href@noop {} {\emph {\bibinfo {title} {Quantum Mechanics:
  Concepts and Applications}}}\ (\bibinfo  {publisher} {Wiley},\ \bibinfo
  {year} {2009})\BibitemShut {NoStop}%
\end{thebibliography}
\end{document}